\documentclass[aps,preprint]{revtex4}
\usepackage{amsfonts}
\usepackage{amsmath}
\usepackage{amssymb}
\usepackage{graphicx}
\usepackage{slashed}

\begin{document}

\title{On Hamiltonian formulations of the Dirac system }
\author{Bence Juh\'{a}sz$^{1}$}
\author{L\'{a}szl\'{o} \'{A}rp\'{a}d Gergely$^{1,2}$}
\date{\today }
\affiliation{$^{1}$ Department of Theoretical Physics, University of Szeged, D\'{o}m t%
\'{e}r 9, H-6720 Szeged, Hungary\\
$^{2}$ Department of Theoretical Physics, HUN-REN Wigner Research Centre for
Physics, Konkoly-Thege Mikl\'{o}s \'{u}t 29-33, H-1121 Budapest, Hungary}

\begin{abstract}
We extend a previously successful discussion of the constrained Schr\"{o}%
dinger system through the Dirac--Bergmann algorithm to the case of the Dirac
field. In order to follow the analogy, first we discuss the classical Dirac
field as a spinorial variable, by introducing properly defined momenta and a
suitably modified, factor ordered Poisson bracket. According to the
Dirac--Bergmann algorithm two second class Hamiltonian constraints emerge,
leading to a factor ordered Dirac bracket on the full phase space. This
becomes the Poisson bracket on the reduced phase space in the canonical
chart adapted to the shell. The Dirac equation is recovered both as
consistency condition on the full phase space and as canonical equation on
the reduced phase space. Alternatively, considering the Dirac field as odd
Grassmann variable, we present the details of the Dirac--Bergmann algorithm
(with either left and right derivatives acting on Grassmann valued
superfunctions and involving a different type of generalized Poisson and
Dirac brackets). We propose a recipe for the canonical second quantization
of all three versions of the generalized Dirac brackets, yielding the
correct fundamental anticommutator.
\end{abstract}

\maketitle

\section{Introduction}

Most of the physical theories of interest contain constraints, for which the
transition from the Lagrangian to the Hamiltonian description requires the
Dirac--Bergmann algorithm, developed initially by Dirac \cite{DiracConstr}
and Bergmann with his collaborators (for a historical account see \cite%
{Salisbury} and for further details \cite{SM,Sundermeyer,HT}). The full
machinery of the formalism through the analysis of a finite-dimensional
system has been recently presented in Ref. \cite{Brown}.

Geometric generalizations of the Dirac--Bergmann algorithm have also been
developed. The algorithm by Gotay, Nester, and Hinds in Ref. \cite{GNH78}
employs a global presymplectic structure, generalizing and improving on the
Dirac--Bergmann algorithm, as the latter can be regarded as its local
version. In the companion paper \cite{GN79}, Gotay and Nester further
emphasize that while the Dirac--Bergmann results cannot always be
transferred to the velocity phase space, their method based on the
presymplectic structure allows that. This was later investigated for a broad
class of admissible Lagrangians, defined as those which allow for
second-order Euler--Lagrange equations on certain submanifolds of the
velocity phase space \cite{GN80}. A multisymplectic approach was also
developed in Ref. \cite{multisymp} and explored in the discussion of various
constrained systems, including Einstein--Hilbert \cite{multisympEH} and
metric-affine \cite{multisympEP} gravity theories, also more recently for
scalar fields, Chern-Simons gravity, and bosonic string theory \cite%
{multisympField}.

While the Dirac--Bergmann algorithm applied for Maxwell electrodynamics or
general relativity \cite{Wald, ADM} are well known, the Schr\"{o}dinger
field as a constrained system was analysed relatively late \cite{Schrod}.
There, the Schr\"{o}dinger equation emerged both as a consistency condition
in the full phase space and as canonical equation in the reduced phase
space. By implementing the quantization scheme for systems with second class
constraints, inconsistencies of previous treatments were eliminated.

In some respects the Dirac field is similar to the Schr\"{o}dinger field.
Most notably, its dynamical equation contains only first time derivatives
(hence a constraint equation), moreover those appear only linearly,
impending the Legendre transformation, and requiring the application of the
Dirac--Bergmann algorithm. The Dirac field, being spinorial, is much more
complicated, yielding anticommutators upon quantization. Hence, it is often
considered an odd Grassmannian variable, in order to recover the same
algebraic structure classically as in the quantum version of the theory.

In this paper, for the classical description of the Dirac field we attempt
to follow as closely as possible the successful treatment of the Schr\"{o}%
dinger field. We will show that the Dirac field classically can be
characterized by a Hamiltonian structure through the algorithmic application
of the Dirac--Bergmann procedure, similarly to the case of the Schr\"{o}%
dinger field. Nevertheless, there is a price to pay due to the Dirac field
being a spinor. In order the Dirac--Bergmann algorithm to work, we will
generalize both the concepts of the Poisson and Dirac brackets such to
observe a meaningful factor ordering. This has to be imposed during the
canonical quantization as well.

The paper proceeds as follows. In Section 2 we introduce the Dirac spinor
(also include a pedagogical discussion on its spin in the Appendix). Then,
we review the variational procedures leading to the Dirac equation in a
classical approach, simplified by the introduction of suitably defined $%
\slashed{\partial}_{a}$\ derivatives.\ The various Lagrangians for the Dirac
field explored in the literature are analogous to the simplest Schiff and
Hermitian Henley--Thirring Lagrangians for the Scr\"{o}dinger field,
discussed in \cite{Schrod}. In contrast with the literature, in our
discussion we define the canonical momenta as Dirac adjoint spinors, which
will have multiple advantages.

Section 3 presents the Dirac--Bergmann algorithm applied to the Hermitian
Lagrangian for the Dirac field. This leads to two primary Hamiltonian
constraints, which are the Dirac adjoints of each other. We proceed with the
definition of the canonical and primary Hamiltonians. Then we introduce a
factor ordered Poisson bracket, in which Dirac adjoint spinors are always
followed by spinors. Time evolution is defined through this Poisson bracket.
The evolution of the constraints generates the Dirac equation\ and its Dirac
adjoint as consistency conditions, holding weakly. The constraints are
second class, which lead to the definition of the Dirac bracket, imposing
again a factor ordering requirement, with the aim to have only Dirac adjoint
spinor times spinor type inner products. Time evolution can be reexpressed
in terms of the factor ordered Dirac bracket. The Dirac brackets of the
canonical pairs contain $1/2$ factors not present in their Poisson brackets.
This feature mimics closely the corresponding feature of the Schr\"{o}dinger
field \cite{Schrod}, which turned advantageous in its canonical
quantization. We identify the reduced phase space variables, proving that
the Dirac bracket on the full phase space turns into the Poisson bracket on
the reduced phase space.\ The Hamiltonian on the reduced phase space
reproduces the Hermitian Hamiltonian evaluated on shell.

For comparing our approach with the interpretation of the classical Dirac
field as an odd Grassmann variable, in Section 4 we also carry out the
Dirac--Bergmann procedure with both left and right derivatives acting on
Grassmann valued superfunctions. These involve another type of generalized
Poisson and Dirac brackets, which are symmetric, rather than antisymmetric,
together with definitions of the momenta different from the literature. As
far as we are aware of, this discussion was not presented in detail
elsewhere.

In Section 5 we proceed with the canonical second quantization of all three
versions of the constrained Dirac dynamics. The suitably defined
quantization recipes transform all versions of Dirac brackets in a natural
way into the fundamental anticommutator of the second quantized Dirac field.

Section 6 contains a brief comparison of our results with the flat limit of
previous discussions of fermionic (Dirac) systems in gravitational field.
Then we present our concluding remarks in Section 7.

We use the metric convention $\eta_{ab}=$diag$\left( -1,1,1,1\right) $.
Latin and Greek indices represent spacetime and space indices, respectively.

\section{Standard variational procedures: a review}

\subsection{The Dirac equation}

The Dirac spinor $\psi $ obeys the equation%
\begin{equation}
\left( i\frac{\hbar }{c}\gamma ^{a}\partial _{a}-m\right) \psi =0~,
\label{DiracEq}
\end{equation}%
where $m$ is the mass and the Dirac matrices $\gamma ^{a}$ are given through
their anticommutators 
\begin{equation}
\left[ \gamma ^{a},\gamma ^{b}\right] _{+}=-2\eta ^{ab}\mathrm{I}_{4}\text{~,%
}  \label{anticommgamma}
\end{equation}%
with $\eta _{ab}~$the Minkowski metric and $\mathrm{I}_{4}$ the unit matrix
in 4 dimensions. In the Dirac representation they read as%
\begin{equation}
\gamma ^{0}=\left( 
\begin{array}{cc}
\mathrm{I}_{2} & 0 \\ 
0 & -\mathrm{I}_{2}%
\end{array}%
\right) ~,\quad \gamma ^{\mu }=\left( 
\begin{array}{cc}
0 & \sigma ^{\mu } \\ 
-\sigma ^{\mu } & 0%
\end{array}%
\right) ~,
\end{equation}%
in the Weyl (chiral) representation as \cite{DWM} 
\begin{equation}
\gamma _{W}^{0}=\left( 
\begin{array}{cc}
0 & \mathrm{I}_{2} \\ 
\mathrm{I}_{2} & 0%
\end{array}%
\right) ~,\quad \gamma _{W}^{\mu }=\left( 
\begin{array}{cc}
0 & -\sigma ^{\mu } \\ 
\sigma ^{\mu } & 0%
\end{array}%
\right) ~,
\end{equation}%
while in the Majorana representation as \cite{DWM}%
\begin{equation}
\gamma _{M}^{0}=\left( 
\begin{array}{cc}
0 & \sigma ^{2} \\ 
\sigma ^{2} & 0%
\end{array}%
\right) ~,\quad \gamma _{M}^{1}=\left( 
\begin{array}{cc}
i\sigma ^{3} & 0 \\ 
0 & i\sigma ^{3}%
\end{array}%
\right) ~,\quad \gamma _{M}^{2}=\left( 
\begin{array}{cc}
0 & -\sigma ^{2} \\ 
\sigma ^{2} & 0%
\end{array}%
\right) ~,\quad \gamma _{M}^{3}=\left( 
\begin{array}{cc}
-i\sigma ^{1} & 0 \\ 
0 & -i\sigma ^{1}%
\end{array}%
\right) ~.
\end{equation}%
(Note that the matrices $\gamma _{W}^{\mu }$ are sometimes defined with a
global minus sign, see Ref. \cite{LancasterBlundell}, which also
equivalently defines $\gamma _{LB}^{1}=-\gamma _{M}^{3}$, $\gamma
_{LB}^{2}=-\gamma _{M}^{2}$, and $\gamma _{LB}^{3}=\gamma _{M}^{1}$.)

In all representations the Dirac matrices are constructed in terms of the
2-dimensional unit matrix $\mathrm{I}_{2}$ and the Pauli matrices $\sigma
^{\mu }$, which obey the commutation relations%
\begin{equation}
\left[ \sigma ^{\mu },\sigma ^{\nu }\right] =2i\epsilon _{~~\rho }^{\mu \nu
}\sigma ^{\rho }~,
\end{equation}%
with $\epsilon _{\mu \nu \rho }$ the Levi-Civita symbol (its indices can
freely raised and lowered due to the metric signature adopted). Explicitly
the Pauli matrices are%
\begin{equation}
\sigma ^{1}=\left( 
\begin{array}{cc}
0 & 1 \\ 
1 & 0%
\end{array}%
\right) ~,\quad \sigma ^{2}=\left( 
\begin{array}{cc}
0 & -i \\ 
i & 0%
\end{array}%
\right) ~,\quad \sigma ^{3}=\left( 
\begin{array}{cc}
1 & 0 \\ 
0 & -1%
\end{array}%
\right) ~,
\end{equation}%
obeying $\left( \sigma ^{\mu }\right) ^{\dag }=\sigma ^{\mu }$ ($^{\dag }$
denotes the adjoint). Hence $\left( \gamma ^{0}\right) ^{\dag }=\gamma ^{0}$
and $\left( \gamma ^{\mu }\right) ^{\dag }=-\gamma ^{\mu }$ (identical
relations hold for all representations). It is also useful to remember the
relations 
\begin{equation}
\left( \gamma ^{0}\right) ^{\dag }\gamma ^{0}=\mathrm{I}_{4}~,\quad \gamma
^{0}\left( \gamma ^{a}\right) ^{\dag }\gamma ^{0}=\gamma ^{a}.  \label{aux}
\end{equation}

In the Appendix we present a simple proof that the Dirac spinor $\psi $
represents spin $1/2$ particles.

\subsection{Lagrangian formalism with $\slashed{\partial}_{a}$ derivatives}

The action 
\begin{equation}
S\left[ \psi ,\bar{\psi}\right] =\int dt\int d\mathbf{r}~\mathfrak{L}\left(
\psi ,\partial _{a}\psi ,\bar{\psi},\partial _{a}\bar{\psi}\right) ~
\end{equation}%
characterizing the Dirac field depends on a pair of independent variables,
the spinor $\psi $ and the Dirac adjoint spinor $\bar{\psi}=\psi ^{\dag
}\gamma ^{0}$. Here $\mathfrak{L}$ is the Lagrangian density, its derivative
with respect to $\psi $ being a Dirac adjoint spinor, while with respect to $%
\bar{\psi}$ a spinor. The Dirac spinor is represented classically as a
column matrix, while its Dirac adjoint spinor as a row matrix. Note that in
order $\mathfrak{L}$\ to be a Lorentz scalar, it has to contain products of
Dirac adjoint spinors with spinors (in this order). Variation of the action
leads to the Euler--Lagrange equations 
\begin{equation}
\frac{\partial \mathfrak{L}}{\partial \psi }-\partial _{a}\frac{\partial 
\mathfrak{L}}{\partial \partial _{a}\psi }=0~,\qquad \frac{\partial 
\mathfrak{L}}{\partial \bar{\psi}}-\partial _{a}\frac{\partial \mathfrak{L}}{%
\partial \partial _{a}\bar{\psi}}=0~.  \label{EL}
\end{equation}%
As will be seen later, all Lagrangians leading to the Dirac equation contain
only Feynman slash (Dirac slash) derivatives $\slashed{\partial}=\gamma
^{a}\partial _{a}$ (with summation over repeated indices understood). We
further introduce the set of derivatives $\slashed{\partial}_{a}$ defined
for spinors and Dirac adjoint spinors through 
\begin{eqnarray}
\slashed{\partial}_{a}\psi &=&\gamma ^{a}\partial _{a}\psi ~~~\left( a!\text{%
, thus no summation over }a\right) ~,  \notag \\
\slashed{\partial}_{a}\bar{\psi} &=&\partial _{a}\bar{\psi}\gamma
^{a}~~~\left( a!\right) ~,  \label{derivatives}
\end{eqnarray}%
respectively. These will be useful for separating temporal and spatial
derivatives, as required in a Hamiltonian treatment. The $\slashed{\partial}%
_{a}$ derivatives exhibit all usual properties of derivative operators,
including the Leibniz rule, also the property%
\begin{equation}
\slashed{\partial}=\sum_{a=0}^{3}\slashed{\partial}_{a}~.
\end{equation}%
Hence all suitable Lagrangians can be rewritten as $\mathfrak{L}\left( \psi ,%
\slashed{\partial}_{a}\psi ,\bar{\psi},\slashed{\partial}_{a}\bar{\psi}%
\right) $. By carrying out partial integrations in the action, the
alternative form of the Euler--Lagrange equations emerge:%
\begin{equation}
\frac{\partial \mathfrak{L}}{\partial \psi }-\slashed{\partial}_{a}\frac{%
\partial \mathfrak{L}}{\partial \slashed{\partial}_{a}\psi }=0~,\qquad \frac{%
\partial \mathfrak{L}}{\partial \bar{\psi}}-\slashed{\partial}_{a}\frac{%
\partial \mathfrak{L}}{\partial \slashed{\partial}_{a}\bar{\psi}}=0~.
\label{ELslash}
\end{equation}%
(The Einstein summation convention applies to the indices of the $%
\slashed{\partial}_{a}$ derivatives.) In detail, they read:%
\begin{equation}
\frac{\partial \mathfrak{L}}{\partial \psi }-\sum_{a=0}^{3}\partial _{a}%
\frac{\partial \mathfrak{L}}{\partial \left( \gamma ^{a}\partial _{a}\psi
\right) }\gamma ^{a}=0~,\qquad \frac{\partial \mathfrak{L}}{\partial \bar{%
\psi}}-\sum_{a=0}^{3}\gamma ^{a}\partial _{a}\frac{\partial \mathfrak{L}}{%
\partial \left( \partial _{a}\bar{\psi}\gamma ^{a}\right) }=0~.
\label{ELslash1}
\end{equation}%
(Due to the triple occurrence of indices, here the summations need to be
explicitly indicated.) As $\left( \gamma ^{0}\right) ^{2}=\mathrm{I}_{4}$
and $\left( \gamma ^{\mu }\right) ^{2}=-\mathrm{I}_{4}$ the Euler--Lagrange
equations (\ref{ELslash1}) formally reduce to (\ref{EL}).

In what follows, we will employ the $\slashed{\partial}_{a}$ derivatives
instead of partial derivatives, as we found them extremely advantageous for
simplifying the formalism. For example, the $\slashed{\partial}_{a}$
derivatives of spinors (or Dirac adjoint spinors) are also spinors (or Dirac
adjoint spinors). In addition, the derivatives of $\mathfrak{L}$ with
respect to $\slashed{\partial}_{a}\psi $ and $\slashed{\partial}_{a}\bar{\psi%
}$ become Dirac adjoint and Dirac spinors, respectively (a property not
shared by the derivatives with respect to $\partial _{a}\psi $ and $\partial
_{a}\bar{\psi}$).

\subsection{The simplest Lagrangian}

The Lagrangian generating the Dirac equation (\ref{DiracEq}) presented in
many quantum field theory textbooks \cite{LancasterBlundell,BD,LL,HK} is%
\begin{equation}
\mathfrak{L}_{BD}=\bar{\psi}\left( i\hbar c\gamma ^{a}\partial
_{a}-mc^{2}\right) \psi ~,  \label{LBD}
\end{equation}%
(the subscript BD refers to the authors of Ref. \cite{BD}, Bjorken and
Drell). Rewriting this with the $\slashed{\partial}$ derivative gives%
\begin{equation}
\mathfrak{L}_{BD}=\bar{\psi}\left( i\hbar c\slashed{\partial}-mc^{2}\right)
\psi ~.  \label{LBDslash}
\end{equation}

Variation of the action with respect to $\bar{\psi}$ (or equivalently,
inserting $\mathfrak{L}_{BD}$ into the Euler--Lagrange equations (\ref%
{ELslash})) immediately gives the Dirac equation (\ref{DiracEq}) in the form$%
~$%
\begin{equation}
\left( i\frac{\hbar }{c}\slashed{\partial}-m\right) \psi =0~,
\label{Diracslash}
\end{equation}%
while varying with respect to $\psi $ leads to%
\begin{equation}
\left( i\frac{\hbar }{c}\slashed{\partial}+m\right) \bar{\psi}=0~,
\label{DiracADslash}
\end{equation}%
which, by virtue of Eqs. (\ref{aux}) is the Dirac adjoint of the Dirac
equation.

The Bjorken--Drell Lagrangian $\mathfrak{L}_{BD}$ is analogue to the one
given by Schiff for the Schr\"{o}dinger field, which has been discussed in
Ref. \cite{Schrod} from a variational point of view. We define the momenta
as 
\begin{align}
\pi ^{BD}& =\frac{\partial \mathfrak{L}_{BD}}{\partial \slashed{\partial}%
_{0}\psi }=i\hbar c\bar{\psi}~,  \notag \\
\bar{\pi}^{BD}& =\frac{\partial \mathfrak{L}_{BD}}{\partial %
\slashed{\partial}_{0}\bar{\psi}}=0~.  \label{momBD}
\end{align}%
This definition differs from the one explored in the literature by
containing the $\slashed{\partial}_{0}$ derivative, resulting in $\pi ^{BD}$
without a $\gamma ^{0}$ multiplier as compared to Ref. \cite{BD}. This
choice however conveniently ensures that $\pi ^{BD}$ is a Dirac adjoint
spinor. Note that momenta with an overbar are spinors and those without are
Dirac adjoint spinors (opposite convention as for the fields).

This modification of their definition does not cure the problem that the
canonical momenta of the Dirac adjoint variables $\psi $ and $\bar{\psi}$
fail to be the Dirac adjoints of each other (as $\bar{\pi}^{BD}$ is
vanishing). A similar disadvantageous property was exhibited in the
canonical treatment of the Schr\"{o}dinger field by Schiff \cite{Schrod}.

Eliminating $\bar{\psi}$ from the action through the first relation (\ref%
{momBD}) leads to the action in already Hamiltonian form%
\begin{equation}
S_{BD}\left[ \psi ,\pi ^{BD}\right] =\int \mathrm{d}t\int \mathrm{d}\mathbf{r%
}~\left( \pi ^{BD}\slashed{\partial}_{0}\psi -\mathcal{H}_{BD}\right) ~,
\label{SBDHam}
\end{equation}%
with the Hamiltonian 
\begin{equation}
\mathcal{H}_{BD}=-\pi ^{BD}\sum_{\mu =1}^{3}\slashed{\partial}_{\mu }\psi -%
\frac{i}{\hbar }mc\left( \pi ^{BD}\psi \right) ~.  \label{HBD}
\end{equation}%
This Hamiltonian is expressed in terms of a single canonical pair of spinors 
$\left( \psi ,\pi ^{BD}\right) $. The canonical equation emerging from the
variation of the functional (\ref{SBDHam}) with respect to $\pi ^{BD}$ gives
the Dirac equation%
\begin{equation}
\frac{\delta S_{BD}\left[ \psi ,\pi ^{BD}\right] }{\delta \pi ^{BD}}=\left( %
\slashed{\partial}+\frac{i}{\hbar }mc\right) \psi =0~,
\end{equation}%
while the variation with respect to $\psi $ yields%
\begin{equation}
\frac{\delta S_{BD}\left[ \psi ,\pi ^{BD}\right] }{\delta \psi }=\left( -%
\slashed{\partial}+\frac{i}{\hbar }mc\right) \pi ^{BD}=0~,
\end{equation}%
which, after reintroducing $\bar{\psi}$ in place of $\pi ^{BD}$ is but the
Dirac adjoint Dirac equation (\ref{DiracADslash}). Hence, there is an
obvious redundancy in this variational approach.

Refs. \cite{LancasterBlundell,BD} advocate for introducing the Dirac
equation into the Hamiltonian $\mathcal{H}_{BD}$ before second quantization,
a procedure yielding%
\begin{equation}
\mathcal{H}_{BD2}=\pi ^{BD}\slashed{\partial}_{0}\psi ~.
\end{equation}%
This however, destroys the variational principle, as it leads to a vanishing
action through Eq. (\ref{SBDHam}), from which no Dirac equation can be
derived.

\subsection{Hermitian Lagrangian}

Instead of the Hamiltonian (\ref{HBD}), expressed in terms of a single
canonical pair of spinors $\left( \psi ,\pi ^{BD}\right) $, another
Hamiltonian of two canonical pairs of complex spinors arises from the
Hermitian Lagrangian \cite{IZ,LL}%
\begin{equation}
\mathfrak{L}_{IZ}=\frac{i\hbar c}{2}\left[ \bar{\psi}\gamma ^{a}\left(
\partial _{a}\psi \right) -\left( \partial _{a}\bar{\psi}\right) \gamma
^{a}\psi \right] -mc^{2}\bar{\psi}\psi  \label{LIZ}
\end{equation}%
(where the subscript IZ refers to Itzykson and Zuber). This is the analogue
of the Henley--Thirring Lagrangian for the Schr\"{o}dinger field, discussed
from a variational point of view in Ref. \cite{Schrod}, which also leads to
a Hamiltonian in terms of two complex fields. The Lagrangian $\mathfrak{L}%
_{IZ}$ arises from $\mathfrak{L}_{BD}$ by subtracting a total divergence $%
\partial _{a}\left[ \left( i\hbar c/2\right) \bar{\psi}\gamma ^{a}\psi %
\right] =\left( i\hbar c/2\right) \slashed{\partial}\left( \bar{\psi}\psi
\right) $.

In terms of the $\slashed{\partial}$ derivative the Lagrangian density reads%
\begin{equation}
\mathfrak{L}_{IZ}=\frac{i\hbar c}{2}\left( \bar{\psi}\slashed{\partial}\psi -%
\slashed{\partial}\bar{\psi}\psi \right) -mc^{2}\bar{\psi}\psi ~.
\label{LIZslash}
\end{equation}%
Variation with respect to the pair of independent spinor variables $\bar{\psi%
}$ and $\psi $ give the Dirac equation (\ref{Diracslash}) and its Dirac
adjoint (\ref{DiracADslash}), respectively. The canonical momenta emerge as%
\begin{align}
\pi & =\frac{\partial \mathfrak{L}_{IZ}}{\partial \slashed{\partial}_{0}\psi 
}=\frac{i\hbar c}{2}\bar{\psi}~,  \notag \\
\bar{\pi}& =\frac{\partial \mathfrak{L}_{IZ}}{\partial \slashed{\partial}_{0}%
\bar{\psi}}=-\frac{i\hbar c}{2}\psi ~.  \label{momIZ}
\end{align}%
They are each other's Dirac adjoint spinors. As before, $\pi $ is a Dirac
adjoint spinor (different by a factor of $\gamma ^{0}$ from the respective
expression in the literature), while $\bar{\pi}$ is a spinor.

Similarly to the case of the Henley--Thirring Lagrangian for the Schr\"{o}%
dinger field \cite{Schrod}, here too we can arrive to an already Hamiltonian
form of the Lagrangian $\mathfrak{L}_{IZ}$ through the following recipe.
First in the kinetic term we replace all $\psi $ and $\bar{\psi}$ appearing
algebraically with their expressions derived from Eqs. (\ref{momIZ}). Then
we separate the mass term into two equal contributions, replacing $\psi $ in
one of the halves and $\bar{\psi}$ in the other one with the respective
expressions containing the momenta. In this way we arrive to 
\begin{equation}
\mathfrak{L}_{IZ}=\pi \slashed{\partial}_{0}\psi +\slashed{\partial}_{0}\bar{%
\psi}\bar{\pi}-\mathcal{H}_{IZ}~.
\end{equation}%
Beside the Liouville form the Hamiltonian density%
\begin{equation}
\mathcal{H}_{IZ}=-\pi \sum_{\mu =1}^{3}\slashed{\partial}_{\mu }\psi
-\sum_{\mu =1}^{3}\slashed{\partial}_{\mu }\bar{\psi}\bar{\pi}-\frac{imc}{%
\hbar }\left( \pi \psi -\bar{\psi}\bar{\pi}\right)  \label{HIZ}
\end{equation}%
emerged. In contrast with the Hamiltonian density (\ref{HBD}) obtained in
the earlier approach, this is Hermitian. Varying the action 
\begin{equation}
S_{IZ}\left[ \psi ,\pi ,\bar{\psi},\bar{\pi}\right] =\int \mathrm{d}t\int 
\mathrm{d}\mathbf{r}~\left[ \pi \slashed{\partial}_{0}\psi +%
\slashed{\partial}_{0}\bar{\psi}\bar{\pi}-\mathcal{H}_{IZ}\right] ~
\end{equation}%
with respect to $\bar{\psi}$ or $\pi $ gives the Dirac equation (\ref%
{Diracslash}), while the variation with respect to $\psi $ or $\bar{\pi}$\
gives its Dirac adjoint (\ref{DiracADslash}):%
\begin{align}
-\frac{2}{c^{2}}\frac{\delta S_{IZ}\left[ \psi ,\pi ,\bar{\psi},\bar{\pi}%
\right] }{\delta \psi }& =\left( i\frac{\hbar }{c}\slashed{\partial}%
+m\right) \bar{\psi}=0~,  \notag \\
i\frac{\hbar }{c}\frac{\delta S_{IZ}\left[ \psi ,\pi ,\bar{\psi},\bar{\pi}%
\right] }{\delta \pi }& =\left( i\frac{\hbar }{c}\slashed{\partial}-m\right)
\psi =0~,  \notag \\
\frac{2}{c^{2}}\frac{\delta S_{IZ}\left[ \psi ,\pi ,\bar{\psi},\bar{\pi}%
\right] }{\delta \bar{\psi}}& =\left( i\frac{\hbar }{c}\slashed{\partial}%
-m\right) \psi =0~,  \notag \\
i\frac{\hbar }{c}\frac{\delta S_{IZ}\left[ \psi ,\pi ,\bar{\psi},\bar{\pi}%
\right] }{\delta \bar{\pi}}& =\left( i\frac{\hbar }{c}\slashed{\partial}%
+m\right) \bar{\psi}=0~.
\end{align}%
Note that all other recipes for interchanging the pair $\left( \psi ,\bar{%
\psi}\right) $ with corresponding expressions of $\left( \pi ,\bar{\pi}%
\right) $ fail to generate the correct Hamiltonian and field equations. This
is consequence of the heavy redundancy in the choice of variables, which
leads to the desired Dirac equation in four equivalent copies.

\section{The Dirac--Bergmann algorithm}

Similarly to the Schr\"{o}dinger field, either of the Lagrangians (\ref%
{LBDslash}) or (\ref{LIZslash}) characterizing the dynamics of the Dirac
spinor contain linearly the generalized velocities $\slashed{\partial}%
_{0}\psi $ and $\slashed{\partial}_{0}\bar{\psi}$, a simptom of a
constrained system. Hence the Dirac--Bergmann algorithm needs to be applied
for an unambiguous transition from the Lagrangian to the Hamiltonian
description.

For this we start from the Hermitian Lagrangian (\ref{LIZslash}). The
generalized momenta (\ref{momIZ}) represent two primary Hamiltonian
constraints%
\begin{align}
\bar{\phi}& :=\pi -\frac{i\hbar c}{2}\bar{\psi}=0~,  \notag \\
\phi & :=\bar{\pi}+\frac{i\hbar c}{2}\psi =0~.  \label{primary}
\end{align}%
The first of this is a constraint on Dirac adjoint spinors, while the second
on spinors and they are the Dirac adjoint of each other.

\subsection{Primary Hamiltonian}

The primary Hamiltonian density arises from the na\"{\i}ve Legendre
transformation%
\begin{equation}
\mathcal{H}_{P}=\pi \left( \slashed{\partial}_{0}\psi \right) +\left( %
\slashed{\partial}_{0}\bar{\psi}\right) \bar{\pi}-\mathfrak{L}_{IZ}~,
\end{equation}%
which can be decomposed as%
\begin{equation}
\mathcal{H}_{P}=\mathcal{H}_{C}+\bar{\phi}\left( \slashed{\partial}_{0}\psi
\right) +\left( \slashed{\partial}_{0}\bar{\psi}\right) \phi ~,  \label{Hp}
\end{equation}%
thus as the sum of the canonical Hamiltonian density 
\begin{equation}
\mathcal{H}_{C}=\frac{i\hbar c}{2}\sum_{\mu =1}^{3}\left[ \left( %
\slashed{\partial}_{\mu }\bar{\psi}\right) \psi -\bar{\psi}\left( %
\slashed{\partial}_{\mu }\psi \right) \right] +mc^{2}\bar{\psi}\psi ~
\label{Hc}
\end{equation}%
with a combination of the primary constraints, in which the spinor $%
\slashed{\partial}_{0}\psi $ and the Dirac adjoint spinor $\slashed{\partial}%
_{0}\bar{\psi}$ are unknown functions of the phase space variables.

\subsection{Factor ordered Poisson bracket}

We introduce \textit{a factor ordered Poisson bracket }(defined by the
requirement that in each term a Dirac adjoint spinor multiplies a spinor) of
two phase space functions $f~$and $g$\ (thus depending only of inner
products of Dirac adjoint spinors with spinors) as%
\begin{eqnarray}
\left\{ f\left( t,\mathbf{r}\right) ,g\left( t,\mathbf{r}^{\prime }\right)
\right\} &=&\int \mathrm{d}\mathbf{r}^{\prime \prime }\left( \frac{\delta
f\left( t,\mathbf{r}\right) }{\delta \psi \left( t,\mathbf{r}^{\prime \prime
}\right) }\frac{\delta g\left( t,\mathbf{r}^{\prime }\right) }{\delta \pi
\left( t,\mathbf{r}^{\prime \prime }\right) }-\frac{\delta g\left( t,\mathbf{%
r}^{\prime }\right) }{\delta \psi \left( t,\mathbf{r}^{\prime \prime
}\right) }\frac{\delta f\left( t,\mathbf{r}\right) }{\delta \pi \left( t,%
\mathbf{r}^{\prime \prime }\right) }\right.  \notag \\
&&\left. +\frac{\delta g\left( t,\mathbf{r}^{\prime }\right) }{\delta \bar{%
\pi}\left( t,\mathbf{r}^{\prime \prime }\right) }\frac{\delta f\left( t,%
\mathbf{r}\right) }{\delta \bar{\psi}\left( t,\mathbf{r}^{\prime \prime
}\right) }-\frac{\delta f\left( t,\mathbf{r}\right) }{\delta \bar{\pi}\left(
t,\mathbf{r}^{\prime \prime }\right) }\frac{\delta g\left( t,\mathbf{r}%
^{\prime }\right) }{\delta \bar{\psi}\left( t,\mathbf{r}^{\prime \prime
}\right) }\right) ~.  \label{PB}
\end{eqnarray}%
The evolution of $f$ is generated by%
\begin{equation}
\slashed{\partial}_{0}f\left( t,\mathbf{r}\right) \approx \left\{ f\left( t,%
\mathbf{r}\right) ,H_{P}\right\} ~,  \label{tevol}
\end{equation}%
where $H_{P}=\int d\mathbf{r}^{\prime }\mathbf{~}\mathcal{H}_{P}\left( t,%
\mathbf{r}^{\prime }\right) $ and $\approx $ stands for weak equality,
holding on the hypersurface defined by the constraints (\ref{primary}).

Note that here the definition of the momenta (\ref{momIZ}) pays off, as it
gets rid of all $\gamma ^{0}$ factors, which would otherwise appear in the
Poisson bracket.

\subsection{Dirac equation as consistency condition}

Eq. (\ref{tevol}) allows to check the consistency of the primary constraints
(\ref{primary}), by imposing that they stay conserved in time:%
\begin{align}
0& \approx \slashed{\partial}_{0}\bar{\phi}\approx \left\{ \bar{\phi}%
,H_{P}\right\} =-\frac{\delta H_{P}}{\delta \psi }-\frac{\delta H_{P}}{%
\delta \bar{\pi}}\frac{i\hbar c}{2}  \notag \\
& \approx -\frac{\delta }{\delta \psi }\int \mathrm{d}\mathbf{r~}\mathcal{H}%
_{C}-\frac{i\hbar c}{2}\frac{\delta }{\delta \bar{\pi}}\int \mathrm{d}%
\mathbf{r~}\mathcal{H}_{C}-i\hbar c\slashed{\partial}_{0}\bar{\psi}  \notag
\\
& =-c^{2}\left( i\frac{\hbar }{c}\slashed{\partial}\bar{\psi}+m\bar{\psi}%
\right)  \label{cons1}
\end{align}%
and%
\begin{align}
0& \approx \slashed{\partial}_{0}\phi \approx \left\{ \phi ,H_{P}\right\} =%
\frac{i\hbar c}{2}\frac{\delta H_{P}}{\delta \pi }-\frac{\delta H_{P}}{%
\delta \bar{\psi}}  \notag \\
& \approx \frac{i\hbar c}{2}\frac{\delta }{\delta \pi }\int \mathrm{d}%
\mathbf{r~}\mathcal{H}_{C}-\frac{\delta }{\delta \bar{\psi}}\int \mathrm{d}%
\mathbf{r~}\mathcal{H}_{C}+i\hbar c\slashed{\partial}_{0}\psi  \notag \\
& =c^{2}\left( i\frac{\hbar }{c}\slashed{\partial}\psi -m\psi \right) ~.
\label{cons2}
\end{align}%
Note, that Eqs. (\ref{cons1})-(\ref{cons2}) represent shorthand notations
for the spinorial components of the constraints $\bar{\phi}$ and $\phi $ (a
Dirac adjoint spinor and a spinor, respectively). This is because the
Poisson bracket was defined on phase space functions.

Remarkably, the Dirac equation\ and its Dirac adjoint both emerged as
consistency conditions, holding weakly. They can also be perceived as the
equations determining the unknown phase space functions $\slashed{\partial}%
_{0}\psi $ and $\slashed{\partial}_{0}\bar{\psi}$.

\subsection{Time derivative}

In the slashed time derivative of a phase space function \thinspace $f$
factor ordering also needs to be imposed:%
\begin{eqnarray}
\slashed{\partial}_{0}f\left( t,\mathbf{r}\right) &=&\int \mathrm{d}\mathbf{r%
}^{\prime \prime }\left( \frac{\delta f\left( t,\mathbf{r}\right) }{\delta
\psi \left( t,\mathbf{r}^{\prime \prime }\right) }\slashed{\partial}_{0}\psi
\left( t,\mathbf{r}^{\prime \prime }\right) +\slashed{\partial}_{0}\bar{\psi}%
\left( t,\mathbf{r}^{\prime \prime }\right) \frac{\delta f\left( t,\mathbf{r}%
\right) }{\delta \bar{\psi}\left( t,\mathbf{r}^{\prime \prime }\right) }%
\right.  \notag \\
&&\left. +\slashed{\partial}_{0}\pi \left( t,\mathbf{r}^{\prime \prime
}\right) \frac{\delta f\left( t,\mathbf{r}\right) }{\delta \pi \left( t,%
\mathbf{r}^{\prime \prime }\right) }+\frac{\delta f\left( t,\mathbf{r}%
\right) }{\delta \bar{\pi}\left( t,\mathbf{r}^{\prime \prime }\right) }%
\slashed{\partial}_{0}\bar{\pi}\left( t,\mathbf{r}^{\prime \prime }\right)
\right) ~.
\end{eqnarray}%
The calculation of the factor ordered Poisson bracket (\ref{PB}) for $%
g=H_{P} $ and exploring Eqs. (\ref{primary}), (\ref{Hp}) and (\ref{Hc})
yields%
\begin{eqnarray}
\left\{ f\left( t,\mathbf{r}\right) ,H_{P}\right\} &=&\slashed{\partial}%
_{0}f\left( t,\mathbf{r}\right) -\int \mathrm{d}\mathbf{r}^{\prime \prime
}\left( i\hbar c\slashed{\partial}+mc^{2}\right) \bar{\psi}\left( t,\mathbf{r%
}^{\prime \prime }\right) \frac{\delta f\left( t,\mathbf{r}\right) }{\delta
\pi \left( t,\mathbf{r}^{\prime \prime }\right) }  \notag \\
&&+\int \mathrm{d}\mathbf{r}^{\prime \prime }\frac{\delta f\left( t,\mathbf{r%
}\right) }{\delta \bar{\pi}\left( t,\mathbf{r}^{\prime \prime }\right) }%
\left( i\hbar c\slashed{\partial}-mc^{2}\right) \psi \left( t,\mathbf{r}%
^{\prime \prime }\right) ~,
\end{eqnarray}%
leading for the phase space variables to the brackets%
\begin{eqnarray}
\left\{ \psi \left( t,\mathbf{r}\right) ,H_{P}\right\} &=&\slashed{\partial}%
_{0}\psi \left( t,\mathbf{r}\right) ~,  \notag \\
\left\{ \bar{\psi}\left( t,\mathbf{r}\right) ,H_{P}\right\} &=&%
\slashed{\partial}_{0}\bar{\psi}\left( t,\mathbf{r}\right) ~,  \notag \\
\left\{ \pi \left( t,\mathbf{r}\right) ,H_{P}\right\} &=&\slashed{\partial}%
_{0}\pi \left( t,\mathbf{r}\right) -\left( i\hbar c\slashed{\partial}%
+mc^{2}\right) \bar{\psi}\left( t,\mathbf{r}\right) ~,  \notag \\
\left\{ \bar{\pi}\left( t,\mathbf{r}\right) ,H_{P}\right\} &=&%
\slashed{\partial}_{0}\bar{\pi}\left( t,\mathbf{r}\right) +\left( i\hbar c%
\slashed{\partial}-mc^{2}\right) \psi \left( t,\mathbf{r}\right) ~.
\end{eqnarray}%
The first two of them are manifestly of the form (\ref{tevol}), while the
third and fourth take the same form when regarded as weak equalities.
Indeed, the consistency conditions of the constraints imply the Dirac
equation and its Dirac adjoint.

\subsection{Factor ordered Dirac bracket}

We can also establish, that the constraints are second class 
\begin{equation}
\left\{ \bar{\phi},\phi \right\} :=\left\{ \bar{\phi}\left( t,\mathbf{r}%
\right) ,\phi \left( t,\mathbf{r}^{\prime }\right) \right\} =-i\hbar c\delta
\left( \mathbf{r}-\mathbf{r}^{\prime }\right) ~.  \label{consBra}
\end{equation}%
This holds component-wise.

In order to rewrite time evolution in terms of a Dirac bracket, we need the
inverse $A^{-1}$ of the distributional matrix 
\begin{equation}
A\left( \mathbf{r}-\mathbf{r}^{\prime }\right) =\left( \left\{ \phi
^{i}\left( t,\mathbf{r}\right) ,\phi ^{j}\left( t,\mathbf{r}^{\prime
}\right) \right\} \right) =\left( 
\begin{array}{cc}
0 & -i\hbar c \\ 
i\hbar c & 0%
\end{array}%
\right) \delta \left( \mathbf{r}-\mathbf{r}^{\prime }\right) ~,
\end{equation}%
where $\phi ^{i}$ stands for $\bar{\phi}$ ($i=1$) and $\phi $ ($i=2$),
defined as\textbf{\ }%
\begin{equation}
\int d\mathbf{r}A^{-1}\left( \mathbf{r}-\mathbf{r}^{\prime \prime }\right)
A\left( \mathbf{r}-\mathbf{r}^{\prime }\right) =\mathrm{I}_{2}\delta \left( 
\mathbf{r}^{\prime }-\mathbf{r}^{\prime \prime }\right) ~,
\end{equation}%
which is\textbf{\ }%
\begin{equation}
A^{-1}\left( \mathbf{r}-\mathbf{r}^{\prime }\right) =\left( 
\begin{array}{cc}
0 & -\frac{i}{\hbar c} \\ 
\frac{i}{\hbar c} & 0%
\end{array}%
\right) \delta \left( \mathbf{r}-\mathbf{r}^{\prime }\right) ~.
\end{equation}%
We also need 
\begin{eqnarray}
\left\{ f\left( t,\mathbf{r}\right) ,\bar{\phi}\left( t,\mathbf{r}^{\prime
}\right) \right\} &=&\frac{\delta f\left( t,\mathbf{r}\right) }{\delta \psi
\left( t,\mathbf{r}^{\prime }\right) }+\frac{i\hbar c}{2}\frac{\delta
f\left( t,\mathbf{r}\right) }{\delta \bar{\pi}\left( t,\mathbf{r}^{\prime
}\right) }~,  \notag \\
\left\{ f\left( t,\mathbf{r}\right) ,\phi \left( t,\mathbf{r}^{\prime
}\right) \right\} &=&-\frac{i\hbar c}{2}\frac{\delta f\left( t,\mathbf{r}%
\right) }{\delta \pi \left( t,\mathbf{r}^{\prime }\right) }+\frac{\delta
f\left( t,\mathbf{r}\right) }{\delta \bar{\psi}\left( t,\mathbf{r}^{\prime
}\right) }~.
\end{eqnarray}%
These are a Dirac adjoint spinor and a spinor, respectively. The Dirac
bracket of two functions $f$, $g$ is defined as 
\begin{gather}
\left\{ f\left( t,\mathbf{r}\right) ,g\left( t,\mathbf{r}^{\prime }\right)
\right\} _{D}=\left\{ f\left( t,\mathbf{r}\right) ,g\left( t,\mathbf{r}%
^{\prime }\right) \right\} \!  \notag \\
-\!\!\int \!\!\!\int \!\mathrm{d}\mathbf{r}^{\prime \prime }\mathrm{d}%
\mathbf{r}^{\prime \prime \prime }\left\{ f\left( t,\mathbf{r}\right) ,\phi
^{i}\left( t,\mathbf{r}^{\prime \prime }\right) \right\} A_{ij}^{-1}\left( 
\mathbf{r}^{\prime \prime }\!-\mathbf{r}^{\prime \prime \prime }\right)
\left\{ \phi ^{j}\left( t,\mathbf{r}^{\prime \prime \prime }\right) ,g\left(
t,\mathbf{r}^{\prime }\right) \right\} ~.
\end{gather}%
In the matrix multiplication \textit{one should pay attention to the factor
ordering, }aiming for Dirac adjoint spinor times spinor type inner products
in each term. By this we mean that the two nonvanishing terms in the sum
have a different factor ordering, resulting in%
\begin{gather}
\left\{ f\left( t,\mathbf{r}\right) ,g\left( t,\mathbf{r}^{\prime }\right)
\right\} _{D}=\left\{ f\left( t,\mathbf{r}\right) ,g\left( t,\mathbf{r}%
^{\prime }\right) \right\}  \notag \\
-\!\!\int \!\!\!\int \!\mathrm{d}\mathbf{r}^{\prime \prime }\mathrm{d}%
\mathbf{r}^{\prime \prime \prime }\left\{ f\left( t,\mathbf{r}\right) ,\bar{%
\phi}\left( t,\mathbf{r}^{\prime \prime }\right) \right\} A_{12}^{-1}\left( 
\mathbf{r}^{\prime \prime }\!-\mathbf{r}^{\prime \prime \prime }\right)
\left\{ \phi \left( t,\mathbf{r}^{\prime \prime \prime }\right) ,g\left( t,%
\mathbf{r}^{\prime }\right) \right\}  \notag \\
-\!\!\int \!\!\!\int \!\mathrm{d}\mathbf{r}^{\prime \prime }\mathrm{d}%
\mathbf{r}^{\prime \prime \prime }\left\{ \bar{\phi}\left( t,\mathbf{r}%
^{\prime \prime \prime }\right) ,g\left( t,\mathbf{r}^{\prime }\right)
\right\} A_{21}^{-1}\left( \mathbf{r}^{\prime \prime }\!-\mathbf{r}^{\prime
\prime \prime }\right) \left\{ f\left( t,\mathbf{r}\right) ,\phi \left( t,%
\mathbf{r}^{\prime \prime }\right) \right\} ~.
\end{gather}%
Calculation gives:%
\begin{gather}
\left\{ f\left( t,\mathbf{r}\right) ,g\left( t,\mathbf{r}^{\prime }\right)
\right\} _{D}=\frac{1}{2}\left\{ f\left( t,\mathbf{r}\right) ,g\left( t,%
\mathbf{r}^{\prime }\right) \right\}  \notag \\
-\frac{i}{\hbar c}\int \mathrm{d}\mathbf{r}^{\prime \prime }\left( \frac{%
\delta f\left( t,\mathbf{r}\right) }{\delta \psi \left( t,\mathbf{r}^{\prime
\prime }\right) }\frac{\delta g\left( t,\mathbf{r}^{\prime }\right) }{\delta 
\bar{\psi}\left( t,\mathbf{r}^{\prime \prime }\right) }-\frac{\delta g\left(
t,\mathbf{r}^{\prime }\right) }{\delta \psi \left( t,\mathbf{r}^{\prime
\prime }\right) }\frac{\delta f\left( t,\mathbf{r}\right) }{\delta \bar{\psi}%
\left( t,\mathbf{r}^{\prime \prime }\right) }\right)  \notag \\
-\frac{i\hbar c}{4}\int \mathrm{d}\mathbf{r}^{\prime \prime }\left( \frac{%
\delta f\left( t,\mathbf{r}\right) }{\delta \bar{\pi}\left( t,\mathbf{r}%
^{\prime \prime }\right) }\frac{\delta g\left( t,\mathbf{r}^{\prime }\right) 
}{\delta \pi \left( t,\mathbf{r}^{\prime \prime }\right) }-\frac{\delta
g\left( t,\mathbf{r}^{\prime }\right) }{\delta \bar{\pi}\left( t,\mathbf{r}%
^{\prime \prime }\right) }\frac{\delta f\left( t,\mathbf{r}\right) }{\delta
\pi \left( t,\mathbf{r}^{\prime \prime }\right) }\right) ~.  \label{DiracBra}
\end{gather}%
Formally, this resembles the Dirac bracket derived for the Schr\"{o}dinger
equation (Eq. (3.7) of Ref. \cite{Schrod}), with the sole (but essential)
difference of factor ordering.

Straightforward calculation shows, that 
\begin{equation}
\slashed{\partial}_{0}f\approx \left\{ f,H_{P}\right\} _{D}~  \label{EqMDB}
\end{equation}%
also holds, as it should for a constrained system without first class
constraints.

\subsection{Dirac brackets of canonical data}

The Dirac brackets (\ref{DiracBra}) of the canonical phase space variables
emerge as%
\begin{equation}
\left\{ \psi \left( t,\mathbf{r}\right) ,\bar{\psi}\left( t,\mathbf{r}%
^{\prime }\right) \right\} _{D}=-\frac{i}{\hbar c}\delta \left( \mathbf{r}-%
\mathbf{r}^{\prime }\right) ~,  \label{canDBa}
\end{equation}%
\begin{equation}
\left\{ \pi \left( t,\mathbf{r}\right) ,\bar{\pi}\left( t,\mathbf{r}^{\prime
}\right) \right\} _{D}=\frac{i\hbar c}{4}\delta \left( \mathbf{r}-\mathbf{r}%
^{\prime }\right) ~,  \label{canDBb}
\end{equation}%
\begin{equation}
\left\{ \psi \left( t,\mathbf{r}\right) ,\psi \left( t,\mathbf{r}^{\prime
}\right) \right\} _{D}=\left\{ \bar{\psi}\left( t,\mathbf{r}\right) ,\bar{%
\psi}\left( t,\mathbf{r}^{\prime }\right) \right\} =0~,  \label{canvan}
\end{equation}%
\begin{equation}
\left\{ \pi \left( t,\mathbf{r}\right) ,\pi \left( t,\mathbf{r}^{\prime
}\right) \right\} _{D}=\left\{ \bar{\pi}\left( t,\mathbf{r}\right) ,\bar{\pi}%
\left( t,\mathbf{r}^{\prime }\right) \right\} =0~,
\end{equation}%
\begin{equation}
\left\{ \psi \left( t,\mathbf{r}\right) ,\bar{\pi}\left( t,\mathbf{r}%
^{\prime }\right) \right\} _{D}=\left\{ \bar{\psi}\left( t,\mathbf{r}\right)
,\pi \left( t,\mathbf{r}^{\prime }\right) \right\} =0~,
\end{equation}%
\begin{equation}
\left\{ \psi \left( t,\mathbf{r}\right) ,\pi \left( t,\mathbf{r}^{\prime
}\right) \right\} _{D}=\frac{1}{2}\left\{ \psi \left( t,\mathbf{r}\right)
,\pi \left( t,\mathbf{r}^{\prime }\right) \right\} =\frac{1}{2}\delta \left( 
\mathbf{r}-\mathbf{r}^{\prime }\right) ~,  \label{canDB1}
\end{equation}%
\begin{equation}
\left\{ \bar{\psi}\left( t,\mathbf{r}\right) ,\bar{\pi}\left( t,\mathbf{r}%
^{\prime }\right) \right\} _{D}=\frac{1}{2}\left\{ \bar{\psi}\left( t,%
\mathbf{r}\right) ,\bar{\pi}\left( t,\mathbf{r}^{\prime }\right) \right\} =%
\frac{1}{2}\delta \left( \mathbf{r}-\mathbf{r}^{\prime }\right) ~.
\label{canDB2}
\end{equation}%
When inserting the definitions (\ref{momIZ}) of the momenta, all nontrivial
Dirac brackets reduce to Eqs. (\ref{canDB1})-(\ref{canDB2}).

\subsection{Reduced phase space}

The Poisson bracket (\ref{consBra}) of the constrains generating a delta
function suggests to introduce the rescaled constraints as a canonical pair
as 
\begin{equation}
\left( 
\begin{array}{c}
\psi _{2}=-\frac{i}{\hbar c}\phi \\ 
\pi _{2}=\bar{\phi}%
\end{array}%
\right) ~.
\end{equation}%
These indeed obey $\left\{ \psi _{2}\left( t,\mathbf{r}\right) ,\pi
_{2}\left( t,\mathbf{r}^{\prime }\right) \right\} =\delta \left( \mathbf{r}-%
\mathbf{r}^{\prime }\right) $. The other pair of canonical variables $\left(
\psi _{1},\pi _{1}\right) $ then emerges by imposing $\left\{ \psi
_{1}\left( t,\mathbf{r}\right) ,\pi _{1}\left( t,\mathbf{r}^{\prime }\right)
\right\} =\delta \left( \mathbf{r}-\mathbf{r}^{\prime }\right) $, $\left\{
\psi _{1}\left( t,\mathbf{r}\right) ,\psi _{2}\left( t,\mathbf{r}^{\prime
}\right) \right\} =0$, $\left\{ \psi _{1}\left( t,\mathbf{r}\right) ,\pi
_{2}\left( t,\mathbf{r}^{\prime }\right) \right\} =0$, $\left\{ \psi
_{2}\left( t,\mathbf{r}\right) ,\pi _{1}\left( t,\mathbf{r}^{\prime }\right)
\right\} =0$ and $\left\{ \pi _{1}\left( t,\mathbf{r}\right) ,\pi _{2}\left(
t,\mathbf{r}^{\prime }\right) \right\} =0$ as%
\begin{equation}
\left( 
\begin{array}{c}
\psi _{1}=\frac{1}{2}\psi +\frac{i}{\hbar c}\bar{\pi} \\ 
\pi _{1}=\frac{i\hbar c}{2}\bar{\psi}+\pi%
\end{array}%
\right) ~.  \label{Psi1Pi1}
\end{equation}%
In the new canonical chart $\psi _{1,2}$ are spinors, while $\pi _{1,2}$ are
Dirac adjoint spinors. The reduced phase space is defined as the surface of
the phase space, where the constraints are obeyed, thus given as $\psi
_{2}=0=\pi _{2}$ and coordinatized by $\left( \psi _{1},\pi _{1}\right) $
only.

Similarly as it happened for the Schr\"{o}dinger case, here too a
straightforward check shows that the Dirac bracket (\ref{DiracBra}) written
in terms of the new coordinates becomes the Poisson bracket of the reduced
phase space. Indeed, under the canonical transformation the Poisson bracket
in the variables $\left( \psi ,\pi ,\bar{\psi},\bar{\pi}\right) $ transforms
to the Poisson bracket in the variables $\left( \psi _{1},\pi _{1},\psi
_{2},\pi _{2}\right) $:%
\begin{equation}
\left\{ f\left( t,\mathbf{r}\right) ,g\left( t,\mathbf{r}^{\prime }\right)
\right\} =\int \mathrm{d}\mathbf{r}^{\prime \prime }\sum_{i=1}^{2}\left( 
\frac{\delta f\left( t,\mathbf{r}\right) }{\delta \psi _{i}\left( t,\mathbf{r%
}^{\prime \prime }\right) }\frac{\delta g\left( t,\mathbf{r}^{\prime
}\right) }{\delta \pi _{i}\left( t,\mathbf{r}^{\prime \prime }\right) }-%
\frac{\delta g\left( t,\mathbf{r}^{\prime }\right) }{\delta \psi _{i}\left(
t,\mathbf{r}^{\prime \prime }\right) }\frac{\delta f\left( t,\mathbf{r}%
\right) }{\delta \pi _{i}\left( t,\mathbf{r}^{\prime \prime }\right) }%
\right) ~.
\end{equation}%
Remarkably, in the new canonical chart there is no need for its factor
ordering, as the Dirac adjoint spinor times spinor structure emerges
naturally.

The complementary terms in the Dirac bracket arising from the second class
constraints give%
\begin{eqnarray}
&&\frac{1}{2}\int \mathrm{d}\mathbf{r}^{\prime \prime }\left( \frac{\delta
f\left( t,\mathbf{r}\right) }{\delta \psi _{1}\left( t,\mathbf{r}^{\prime
\prime }\right) }\frac{\delta g\left( t,\mathbf{r}^{\prime }\right) }{\delta
\pi _{1}\left( t,\mathbf{r}^{\prime \prime }\right) }-\frac{\delta g\left( t,%
\mathbf{r}^{\prime }\right) }{\delta \psi _{1}\left( t,\mathbf{r}^{\prime
\prime }\right) }\frac{\delta f\left( t,\mathbf{r}\right) }{\delta \pi
_{1}\left( t,\mathbf{r}^{\prime \prime }\right) }\right)   \notag \\
&&-\frac{1}{2}\int \mathrm{d}\mathbf{r}^{\prime \prime }\left( \frac{\delta
f\left( t,\mathbf{r}\right) }{\delta \psi _{2}\left( t,\mathbf{r}^{\prime
\prime }\right) }\frac{\delta g\left( t,\mathbf{r}^{\prime }\right) }{\delta
\pi _{2}\left( t,\mathbf{r}^{\prime \prime }\right) }-\frac{\delta g\left( t,%
\mathbf{r}^{\prime }\right) }{\delta \psi _{2}\left( t,\mathbf{r}^{\prime
\prime }\right) }\frac{\delta f\left( t,\mathbf{r}\right) }{\delta \pi
_{2}\left( t,\mathbf{r}^{\prime \prime }\right) }\right) ~.
\end{eqnarray}%
Thus the Dirac bracket becomes%
\begin{equation}
\left\{ f\left( t,\mathbf{r}\right) ,g\left( t,\mathbf{r}^{\prime }\right)
\right\} _{D}=\int \mathrm{d}\mathbf{r}^{\prime \prime }\left( \frac{\delta
f\left( t,\mathbf{r}\right) }{\delta \psi _{1}\left( t,\mathbf{r}^{\prime
\prime }\right) }\frac{\delta g\left( t,\mathbf{r}^{\prime }\right) }{\delta
\pi _{1}\left( t,\mathbf{r}^{\prime \prime }\right) }-\frac{\delta g\left( t,%
\mathbf{r}^{\prime }\right) }{\delta \psi _{1}\left( t,\mathbf{r}^{\prime
\prime }\right) }\frac{\delta f\left( t,\mathbf{r}\right) }{\delta \pi
_{1}\left( t,\mathbf{r}^{\prime \prime }\right) }\right) ~,  \label{DBred}
\end{equation}%
which is exactly the Poisson bracket on the reduced phase space. It is
straightforward to verify the generic property that the Dirac brackets of
the constraints $\psi _{2}$ and $\pi _{2}$ with arbitrary functions vanish.

The time evolutions are still given as%
\begin{equation}
\slashed{\partial}_{0}f\approx \left\{ f,H_{P}\right\} _{D}~.
\end{equation}%
To prove this, we note that the primary Hamiltonian density in the new
canonical chart becomes 
\begin{eqnarray}
\mathcal{H}_{P} &=&\frac{1}{2}\left[ \slashed{\partial}\left( \pi _{1}-\pi
_{2}\right) \right] \psi _{2}+\frac{1}{2}\pi _{2}\slashed{\partial}\left(
\psi _{1}+\psi _{2}\right)  \notag \\
&&+\frac{1}{2}\sum_{\mu =1}^{3}\left[ \slashed{\partial}_{\mu }\left( \pi
_{1}-\pi _{2}\right) \right] \psi _{1}+\frac{1}{2}\left[ \slashed{\partial}%
_{0}\left( \pi _{1}-\pi _{2}\right) \right] \psi _{2}  \notag \\
&&-\frac{1}{2}\pi _{1}\sum_{\mu =1}^{3}\slashed{\partial}_{\mu }\left( \psi
_{1}+\psi _{2}\right) +\frac{1}{2}\pi _{2}\slashed{\partial}_{0}\left( \psi
_{1}+\psi _{2}\right)  \notag \\
&&-\frac{ic}{\hbar }m\left( \pi _{1}-\pi _{2}\right) \left( \psi _{1}+\psi
_{2}\right) ~.
\end{eqnarray}%
Then, 
\begin{eqnarray}
\left\{ f,H_{P}\right\} _{D}~ &=&\int \mathrm{d}\mathbf{r}^{\prime \prime }%
\left[ \frac{\delta f\left( t,\mathbf{r}\right) }{\delta \psi _{1}\left( t,%
\mathbf{r}^{\prime \prime }\right) }\slashed{\partial}_{0}\psi _{1}\left( t,%
\mathbf{r}^{\prime \prime }\right) +\slashed{\partial}_{0}\pi _{1}\left( t,%
\mathbf{r}^{\prime \prime }\right) \frac{\delta f\left( t,\mathbf{r}\right) 
}{\delta \pi _{1}\left( t,\mathbf{r}^{\prime \prime }\right) }\right.  \notag
\\
&&+\frac{\delta f\left( t,\mathbf{r}\right) }{\delta \psi _{1}\left( t,%
\mathbf{r}^{\prime \prime }\right) }\frac{ic}{\hbar }\left( i\frac{\hbar }{c}%
\slashed{\partial}-m\right) \left( \psi _{1}+\psi _{2}\right) \left( t,%
\mathbf{r}^{\prime \prime }\right)  \notag \\
&&\left. +\frac{ic}{\hbar }\left( i\frac{\hbar }{c}\slashed{\partial}%
+m\right) \left( \pi _{1}-\pi _{2}\right) \left( t,\mathbf{r}^{\prime \prime
}\right) \frac{\delta f\left( t,\mathbf{r}\right) }{\delta \pi _{1}\left( t,%
\mathbf{r}^{\prime \prime }\right) }\right] ~.
\end{eqnarray}%
On the constraint surface the first line is but the slashed time derivative,
as the consistency conditions can be rewritten as $\slashed{\partial}%
_{0}\psi _{2}=0=\slashed{\partial}_{0}\pi _{2}$, allowing to add the two
missing terms to the total time derivative. The second and third lines also
vanish on shell as $\psi _{1}+\psi _{2}=\psi $ and $\pi _{1}-\pi _{2}=i\hbar
c\bar{\psi}$ (which obey the Dirac equation and its Dirac adjoint, as
consistency conditions).

We conclude the subsection by noting that the Hamiltonian $\mathcal{H}_{R}$
on the reduced phase space, obtained by rewriting $\mathcal{H}_{P}\approx 
\mathcal{H}_{C}$ in the canonical chart $\left( \psi _{1},\pi _{1},\psi
_{2}=0,\pi _{2}=0\right) $ can be expressed in the alternative forms%
\begin{eqnarray}
\mathcal{H}_{R} &=&\frac{1}{2}\sum_{\mu =1}^{3}\left[ \left( %
\slashed{\partial}_{\mu }\pi _{1}\right) \psi _{1}-\pi _{1}\left( %
\slashed{\partial}_{\mu }\psi _{1}\right) \right] -\frac{i}{\hbar }mc\pi
_{1}\psi _{1}  \notag \\
&=&\sum_{\mu =1}^{3}\slashed{\partial}_{\mu }\left( \frac{\pi _{1}\psi _{1}}{%
2}\right) +\mathcal{H}_{BD}\left( \psi _{1},\pi _{1}\right)  \notag \\
&=&\mathcal{H}_{IZ}\left( \psi _{1},\pi _{1},\psi _{2}=0,\pi _{2}=0\right) ~.
\label{HR}
\end{eqnarray}%
This is slightly different from the Schr\"{o}dinger case, where the
Hamiltonian on the reduced phase space coincides with both the Schiff and
Henley--Thirring Hamiltonians, when evaluated on shell.

Variation of the action 
\begin{equation}
S_{R}=\int \mathrm{d}^{4}x\left[ \pi _{1}\slashed{\partial}_{0}\psi _{1}-%
\mathcal{H}_{R}\right]
\end{equation}%
with respect to the reduced space variables $\psi _{1}$ and $\pi _{1}$ gives%
\begin{align}
\frac{\delta S_{R}\left[ \psi _{1},\pi _{1}\right] }{\delta \psi _{1}}& =i%
\frac{c}{\hbar }\left( i\frac{\hbar }{c}\slashed{\partial}+m\right) \pi
_{1}=0~,  \notag \\
\frac{\delta S_{R}\left[ \psi _{1},\pi _{1}\right] }{\delta \pi _{1}}& =-i%
\frac{c}{\hbar }\left( i\frac{\hbar }{c}\slashed{\partial}-m\right) \psi
_{1}=0~.
\end{align}%
These are the Dirac equation and its Dirac adjoint, respectively, both
obtained as canonical equations on the reduced phase space.

Hence, at a classical level the Dirac field can be suitably characterized
through a Hamiltonian evolution through the Dirac--Bergmann method, in full
analogy with the corresponding procedure for the Schr\"{o}dinger field. The
price to pay was the need for introducing factor ordered Poisson and Dirac
brackets. The discussion was simplified by the modified definitions of the
momenta as compared to the standard literature. Then we gave the Dirac
brackets of the canonical data as Eqs. (\ref{canvan})-(\ref{canDB2}). We
also identified the reduced phase space and the reduced Hamiltonian (\ref{HR}%
) on it.

\section{The Dirac field as Grassmann variable}

In an alternative approach explored in Refs. \cite{HT} and \cite{Grassmann2}%
, classically fermionic fields are regarded as odd Grassmann variables,
nevertheless these references do not present the case of the Dirac field
explicitly. For comparing with our earlier treatment, we explicitly present
the Dirac--Bergmann algorithm for the Dirac field, treated as an odd
Grassmann variable, in two equivalent formulations. For simplifying the
notation, we also drop the arguments of the variables in this section.

\subsection{Grassmann variables}

The generators $\xi ^{i}$ of a Grassmann algebra anticommute as 
\begin{equation}
\xi ^{i}\xi ^{j}=-\xi ^{j}\xi ^{i}~.
\end{equation}%
In the above notation we have suppressed the exterior or wedge product
between them. Such products applied repeatedly on the generators lead to
superfunctions. By definition, a subclass of superfunctions anticommuting
with each other are odd (of parity $1$). Another subclass, consisting of
superfunctions commuting with each other and also with the odd ones are even
(of parity $0$). The simplest superfunctions are the odd (even) Grassmann
variables, also dubbed Grassmann numbers, defined as series of odd (even)
number products of generators with complex coefficients. Classically the
Dirac spinor, its adjoint and Dirac adjoint all qualify as odd Grassman
variables, which anticommute, for example $\bar{\psi}\psi =-\psi \bar{\psi}$%
. By contrast, an even Grassmann variable $E$ commutes with both even or odd
Grassman variables, or in fact with any superfunction $F$ as $EF=FE$.
Following this approach there is no need for factor ordering, as $\psi ,\
\psi ^{\dag },\ \Bar{\psi}$ appear on equal footing (they can be thought of
as $4\times 4$ matrices). The price to pay is that when calculating the
brackets of the superfunctions, two versions of partial derivatives with
respect to the generators need to be introduced. These are the left (denoted 
$L$, acting from the left to the right) and right ($R$, acting from the
right to the left) derivatives, related as 
\begin{equation}
\dfrac{\partial ^{L}F}{\partial \xi ^{i}}=(-1)^{1+\epsilon _{F}}\dfrac{%
\partial ^{R}F}{\partial \xi ^{i}}~,  \label{RL}
\end{equation}%
where $\epsilon _{F}$ indicates the parity of $F$. We illustrate the action
of these derivatives on the product $\xi ^{1}\xi ^{2}$ as follows%
\begin{eqnarray}
\dfrac{\partial ^{L}}{\partial \xi ^{1}}\left( \xi ^{1}\xi ^{2}\right)
&=&\xi ^{2}~,\quad \dfrac{\partial ^{L}}{\partial \xi ^{2}}\left( \xi
^{1}\xi ^{2}\right) =-\xi ^{1}~,  \notag \\
\dfrac{\partial ^{R}}{\partial \xi ^{1}}\left( \xi ^{1}\xi ^{2}\right)
&=&-\xi ^{2}~,\quad \dfrac{\partial ^{R}}{\partial \xi ^{2}}\left( \xi
^{1}\xi ^{2}\right) =\xi ^{1}~.
\end{eqnarray}%
In general, the action of a left derivative on a product of any Grassmann
superfunctions $FG$ with respect to an odd Grassmann variable $\psi $ is 
\cite{Berezin}:%
\begin{equation}
\dfrac{\partial ^{L}\left( FG\right) }{\text{ }\partial \psi }=\dfrac{%
\partial ^{L}F}{\text{ }\partial \psi }G+(-1)^{\epsilon _{F}}F\dfrac{%
\partial ^{L}G}{\text{ }\partial \psi }~,
\end{equation}%
whereas a right derivative acts as%
\begin{equation}
\dfrac{\partial ^{R}\left( FG\right) }{\text{ }\partial \psi }=F\dfrac{%
\partial ^{R}G}{\text{ }\partial \psi }+(-1)^{\epsilon _{G}}\dfrac{\partial
^{R}F}{\text{ }\partial \psi }G~.
\end{equation}%
Similar rules apply for differentiation with respect to $\bar{\psi}$. For
the derivative with respect to an even Grassmann variable the Leibniz rule
applies for both types of derivatives.

\subsection{Hamiltonian formalism in terms of left derivatives}

For the Lagrangian of the Dirac system we once again adopt (\ref{LIZslash}),
denoted $\mathfrak{L}_{G}$ to emphasize that this time it contains Grassmann
variables, hence it is a superdensity, of parity $0$. As before, we define
the momenta as derivatives with respect to the $\slashed{\partial}_{0}$
derivatives of the field and its Dirac adjoint. However it has to be
specified whether the derivative is left or right. While both conventions
work, in this subsection we follow Ref. \cite{HT}, in adopting the left
derivative. We define the general momenta in terms of left derivatives as:%
\begin{eqnarray}
\pi _{L} &=&\dfrac{\partial ^{L}\mathfrak{L}_{G}}{\partial \slashed{\partial}%
_{0}\psi }=-\dfrac{i\hbar c}{2}\Bar{\psi}~,  \notag \\
\Bar{\pi}_{L} &=&\dfrac{\partial ^{L}\mathfrak{L}_{G}}{\partial %
\slashed{\partial}_{0}\Bar{\psi}}=-\dfrac{i\hbar c}{2}\psi ~.  \label{defL}
\end{eqnarray}%
They differ by $\gamma ^{0}$ factors in the denominator from the expressions
obtainable by applying the definitions advanced in Ref. \cite{HT} (which
contain $\partial _{0}$ instead $\slashed{\partial}_{0}$). The purpose of
this change is to remain as close as possible to the expressions of the
momenta introduced in the previous chapter, Eqs. (\ref{momIZ}). Indeed, $%
\bar{\pi}_{L}$ has the same form as $\bar{\pi}$, however $\pi _{L}$ differs
in a minus sign. The momenta are odd variables.

Following the same procedure as before (also paying attention that when
employing left derivatives, in order to reproduce the momenta, the Liouville
form needs to be defined as $\slashed{\partial}_{0}\psi \pi _{L}+%
\slashed{\partial}_{0}\Bar{\psi}\Bar{\pi}_{L}$), the Hamilton superdensity
emerges:%
\begin{equation}
\mathcal{H}_{L}=-\sum_{\mu }\slashed{\partial}_{\mu }\psi \pi _{L}-\sum_{\mu
}\slashed{\partial}_{\mu }\Bar{\psi}\Bar{\pi}_{L}+\dfrac{imc}{\hbar }(\psi
\pi _{L}+\Bar{\psi}\Bar{\pi}_{L})~.  \label{Hamilton left}
\end{equation}%
The Hamiltonian equations are%
\begin{eqnarray}
\slashed{\partial}_{0}\psi &=&-\dfrac{\partial ^{L}\mathcal{H}_{L}}{\partial
\pi _{L}}~,\quad \slashed{\partial}_{0}\Bar{\psi}=-\dfrac{\partial ^{L}%
\mathcal{H}_{L}}{\partial \Bar{\pi}_{L}}~,  \notag \\
\slashed{\partial}_{0}\pi _{L} &=&-\dfrac{\partial ^{L}\mathcal{H}_{L}}{%
\partial \psi }~,\quad \slashed{\partial}_{0}\Bar{\pi}_{L}=-\dfrac{\partial
^{L}\mathcal{H}_{L}}{\partial \Bar{\psi}}~.
\end{eqnarray}%
After accounting for the change in the definitions of momenta, these agree
with the expressions obtainable through the canonical equations given in
Ref. \cite{HT}. Note that the left derivatives above can also be rewritten
as right derivatives through Eq. (\ref{RL}).

The generalized Poisson bracket of generic superfunctions $F,G$ with respect
to canonical data of the Dirac field is: 
\begin{equation}
\{F,G\}_{+}^{L}=\int \mathrm{d}\mathbf{r}\Bigg[(-1)^{\epsilon _{F}}\bigg(%
\dfrac{\delta ^{L}F}{\delta \psi }\dfrac{\delta ^{L}G}{\delta \pi _{L}}+%
\dfrac{\delta ^{L}F}{\delta \pi _{L}}\dfrac{\delta ^{L}G}{\delta \psi }+%
\dfrac{\delta ^{L}F}{\delta \Bar{\psi}}\dfrac{\delta ^{L}G}{\delta \Bar{\pi}%
_{L}}+\dfrac{\delta ^{L}F}{\delta \Bar{\pi}_{L}}\dfrac{\delta ^{L}G}{\delta 
\Bar{\psi}}\bigg)\Bigg]~.  \label{PL}
\end{equation}%
It obeys 
\begin{equation}
\{F,G\}_{+}^{L}=-(-1)^{\epsilon _{F}\epsilon _{G}}\{G,F\}_{+}^{L}~.
\label{Pprop}
\end{equation}%
The Poisson brackets of canonical variables become:%
\begin{eqnarray}
\{\psi ,\Bar{\psi}\}_{+}^{L} &=&\{\Bar{\psi},\psi \}_{+}^{L}=\{\pi _{L},\Bar{%
\pi}_{L}\}_{+}^{L}=\{\Bar{\pi}_{L},\pi _{L}\}_{+}^{L}=0~,  \notag \\
\{\psi ,\pi _{L}\}_{+}^{L} &=&\{\pi _{L},\psi \}_{+}^{L}=\{\Bar{\psi},\Bar{%
\pi}_{L}\}_{+}^{L}=\{\bar{\pi}_{L},\Bar{\psi}\}_{+}^{L}=-\delta ({%
\boldsymbol{r}}-{\boldsymbol{r}}^{\prime })~.
\end{eqnarray}%
These Poisson brackets are symmetric.

\subsection{Hamiltonian formalism in terms of right derivatives}

The dynamics can equally be expressed in terms of right derivatives, as
follows. The momenta in this case are conveniently defined as%
\begin{eqnarray}
\pi _{R} &=&\dfrac{\partial ^{R}\mathfrak{L}_{G}}{\partial \slashed{\partial}%
_{0}\psi }=\dfrac{i\hbar c}{2}\Bar{\psi}~,  \notag \\
\Bar{\pi}_{R} &=&\dfrac{\partial ^{R}\mathfrak{L}_{G}}{\partial %
\slashed{\partial}_{0}\Bar{\psi}}=\dfrac{i\hbar c}{2}\psi ~.  \label{defR}
\end{eqnarray}%
Note the differences in signs as compared with $\pi _{L}$ and $\bar{\pi}_{L}$%
.

The Liouville form now becomes $\pi _{R}\slashed{\partial}_{0}\psi +\Bar{\pi}%
_{R}\slashed{\partial}_{0}\Bar{\psi}$ in order to be in agreement with the
definitions (\ref{defR}), leading to the Hamiltonian superdensity: 
\begin{equation}
\mathcal{H}_{R}=-\pi _{R}\sum_{\mu }\slashed{\partial}_{\mu }\psi -\Bar{\pi}%
_{R}\sum_{\mu }\slashed{\partial}_{\mu }\Bar{\psi}+\dfrac{imc}{\hbar }(\pi
_{R}\psi +\Bar{\pi}_{R}\Bar{\psi})~,  \label{Hamilton right}
\end{equation}%
an expression different from (\ref{Hamilton left}) in its factor ordering.
The Hamiltonian equations read 
\begin{eqnarray}
\slashed{\partial}_{0}\psi &=&-\dfrac{\partial ^{R}\mathcal{H}_{R}}{\partial
\pi _{R}}~,\quad \slashed{\partial}_{0}\Bar{\psi}=-\dfrac{\partial ^{R}%
\mathcal{H}_{R}}{\partial \Bar{\pi}_{R}}~,  \notag \\
\slashed{\partial}_{0}\pi _{R} &=&-\dfrac{\partial ^{R}\mathcal{H}_{R}}{%
\partial \psi }~,\quad \slashed{\partial}_{0}\Bar{\pi}_{R}=-\dfrac{\partial
^{R}\mathcal{H}_{R}}{\partial \Bar{\psi}}~.
\end{eqnarray}%
After accounting for the change in the definitions of momenta, these agree
with the corresponding equations presented in Ref. \cite{Grassmann2}. They
can also be rewritten in terms of left derivatives, if needed.

The generalized Poisson bracket with right derivatives is:%
\begin{equation}
\{F,G\}_{+}^{R}=\int \mathrm{d}\mathbf{r}\Bigg[(-1)^{\epsilon _{G}}\bigg(%
\dfrac{\delta ^{R}F}{\delta \psi }\dfrac{\delta ^{R}G}{\delta \pi _{R}}+%
\dfrac{\delta ^{R}F}{\delta \pi _{R}}\dfrac{\delta ^{R}G}{\delta \psi }+%
\dfrac{\delta ^{R}F}{\delta \Bar{\psi}}\dfrac{\delta ^{R}G}{\delta \Bar{\pi}%
_{R}}+\dfrac{\delta ^{R}F}{\delta \Bar{\pi}_{R}}\dfrac{\delta ^{R}G}{\delta 
\Bar{\psi}}\bigg)\Bigg].  \label{PR}
\end{equation}%
It also obeys a similar property to (\ref{Pprop}). Finally, the Poisson
brackets of canonical variables are%
\begin{eqnarray}
\{\psi ,\Bar{\psi}\}_{+}^{R} &=&\{\Bar{\psi},\psi \}_{+}^{R}=\{\pi _{R},\Bar{%
\pi}_{R}\}_{+}^{R}=\{\Bar{\pi}_{R},\pi _{R}\}_{+}^{R}=0~,  \notag \\
\{\psi ,\pi _{R}\}_{+}^{R} &=&\{\pi _{R},\psi \}_{+}^{R}=\{\Bar{\psi},\Bar{%
\pi}_{R}\}_{+}^{R}=\{\bar{\pi}_{R},\Bar{\psi}\}_{+}^{R}=-\delta ({%
\boldsymbol{r}}-{\boldsymbol{r}}^{\prime })~.
\end{eqnarray}%
We remark that despite the different brackets employed here and in the
previous subsection, the two algebras are identical.

\subsection{Dirac-Bergmann algorithm}

For brevity we present the Dirac--Bergmann algorithm in terms of the two
types of derivatives in parallel. The constrains are 
\begin{equation}
\phi ^{L}=\Bar{\pi}_{L}+\dfrac{i\hbar c}{2}\psi ~,\quad \Bar{\phi}^{L}=\pi
_{L}+\dfrac{i\hbar c}{2}\Bar{\psi}~,  \label{primaryL}
\end{equation}%
\begin{equation}
\phi ^{R}=\Bar{\pi}_{R}-\dfrac{i\hbar c}{2}\psi ~,\quad \Bar{\phi}^{R}=\pi
_{R}-\dfrac{i\hbar c}{2}\Bar{\psi}~.  \label{primaryR}
\end{equation}%
The primary Hamiltonian superdensities follow: 
\begin{eqnarray}
\mathcal{H}_{PL} &=&\mathcal{H}_{L}+\slashed{\partial}_{0}\psi \Bar{\phi}%
^{L}+\slashed{\partial}_{0}\Bar{\psi}\phi ^{L}  \notag \\
&=&-\sum_{\mu }\slashed{\partial}_{\mu }\psi \pi _{L}-\sum_{\mu }%
\slashed{\partial}_{\mu }\Bar{\psi}\Bar{\pi}_{L}+\dfrac{imc}{\hbar }(\psi
\pi _{L}+\Bar{\psi}\Bar{\pi}_{L})+\slashed{\partial}_{0}\psi \Bar{\phi}^{L}+%
\slashed{\partial}_{0}\Bar{\psi}\phi ^{L}~,
\end{eqnarray}%
\begin{eqnarray}
\mathcal{H}_{PR} &=&\mathcal{H}_{R}+\Bar{\phi}^{R}\slashed{\partial}_{0}\psi
+\phi ^{R}\slashed{\partial}_{0}\Bar{\psi}  \notag \\
&=&-\pi _{R}\sum_{\mu }\slashed{\partial}_{\mu }\psi -\Bar{\pi}_{R}\sum_{\mu
}\slashed{\partial}_{\mu }\Bar{\psi}+\dfrac{imc}{\hbar }(\pi _{R}\psi +\Bar{%
\pi}_{R}\Bar{\psi})+\Bar{\phi}^{R}\slashed{\partial}_{0}\psi +\phi ^{R}%
\slashed{\partial}_{0}\Bar{\psi}~.
\end{eqnarray}%
As before, the time evolutions%
\begin{equation}
0\approx \slashed{\partial}_{0}\phi =\{\phi ,H_{PL,R}\}_{+}^{L,R}
\end{equation}%
conserve the constraints (where $\phi $ stands for any of $\phi ^{L,R},~\bar{%
\phi}^{L,R}$).

The Poisson brackets of constrains: 
\begin{equation}
\{\Bar{\phi}^{L},\phi ^{L}\}_{+}^{L}=-\int \mathrm{d}\mathbf{r}\bigg [\dfrac{%
\delta ^{L}\Bar{\phi}^{L}}{\delta \pi _{L}}\dfrac{\delta ^{L}\phi ^{L}}{%
\delta \psi }+\dfrac{\delta ^{L}\Bar{\phi}^{L}}{\delta \Bar{\psi}}\dfrac{%
\delta ^{L}\phi ^{L}}{\delta \Bar{\pi}_{L}}\bigg]=-i\hbar c\delta ({%
\boldsymbol{r}}-{\boldsymbol{r}}^{\prime })\ ,
\end{equation}%
\begin{equation}
\{\Bar{\phi}^{R},\phi ^{R}\}_{+}^{R}=-\int \mathrm{d}\mathbf{r}\bigg [\dfrac{%
\delta ^{R}\Bar{\phi}^{R}}{\delta \pi _{R}}\dfrac{\delta ^{R}\phi ^{R}}{%
\delta \psi }+\dfrac{\delta ^{R}\Bar{\phi}^{R}}{\delta \Bar{\psi}}\dfrac{%
\delta ^{R}\phi ^{R}}{\delta \Bar{\pi}_{R}}\bigg]=i\hbar c\delta ({%
\boldsymbol{r}}-{\boldsymbol{r}}^{\prime })
\end{equation}%
generate the matrices%
\begin{equation}
A_{ij}^{L}=-A_{ij}^{R}=%
\begin{bmatrix}
0 & -i\hbar c \\ 
-i\hbar c & 0%
\end{bmatrix}%
\delta ({\boldsymbol{r}}-{\boldsymbol{r}}^{\prime })~,\qquad
(A_{ij}^{L})^{-1}=-(A_{ij}^{R})^{-1}=%
\begin{bmatrix}
0 & \frac{i}{\hbar c} \\ 
\frac{i}{\hbar c} & 0%
\end{bmatrix}%
\delta ({\boldsymbol{r}}-{\boldsymbol{r}}^{\prime })~,
\end{equation}%
allowing to define the Dirac brackets through 
\begin{equation}
\{F,G\}_{D+}^{L,R}=\{F,G\}_{+}^{L,R}-\int \int \mathrm{d}\mathbf{r}^{\prime
\prime }\mathrm{d}\mathbf{r}^{\prime \prime \prime }\{f,\phi
^{i}\}_{+}^{L,R}(A_{ij}^{L,R})^{-1}\{\phi ^{j},g\}_{+}^{L,R}~.
\end{equation}%
As a result, the following two Dirac brackets emerge 
\begin{align}
\{F,G\}_{D+}^{L}=\dfrac{1}{2}\{F,G\}_{+}^{L}& +(-1)^{\epsilon _{f}}\dfrac{i}{%
\hbar c}\int \mathrm{d}\mathbf{r}\bigg[\dfrac{\delta ^{L}F}{\delta \psi }%
\dfrac{\delta ^{L}G}{\delta \Bar{\psi}}+\dfrac{\delta ^{L}F}{\delta \Bar{\psi%
}}\dfrac{\delta ^{L}G}{\delta \psi }\bigg]  \notag \\
& -(-1)^{\epsilon _{F}}\dfrac{i\hbar c}{4}\int \mathrm{d}\mathbf{r}\bigg[%
\dfrac{\delta ^{L}F}{\delta \pi _{L}}\dfrac{\delta ^{L}G}{\delta \Bar{\pi}%
_{L}}+\dfrac{\delta ^{L}F}{\delta \Bar{\pi}_{L}}\dfrac{\delta ^{L}G}{\delta
\pi _{L}}\bigg]\ ,  \label{DL}
\end{align}%
\begin{align}
\{F,G\}_{D+}^{R}=\dfrac{1}{2}\{F,G\}_{+}^{R}& -(-1)^{\epsilon _{G}}\dfrac{i}{%
\hbar c}\int \mathrm{d}\mathbf{r}\bigg[\dfrac{\delta ^{R}F}{\delta \psi }%
\dfrac{\delta ^{R}G}{\delta \Bar{\psi}}+\dfrac{\delta ^{R}F}{\delta \Bar{\psi%
}}\dfrac{\delta ^{R}G}{\delta \psi }\bigg]  \notag \\
& +(-1)^{\epsilon _{G}}\dfrac{i\hbar c}{4}\int \mathrm{d}\mathbf{r}\bigg[%
\dfrac{\delta ^{R}F}{\delta \pi _{R}}\dfrac{\delta ^{R}G}{\delta \Bar{\pi}%
_{R}}+\dfrac{\delta ^{R}F}{\delta \Bar{\pi}_{R}}\dfrac{\delta ^{R}G}{\delta
\pi _{R}}\bigg]\ .  \label{DR}
\end{align}%
In particular, the nontrivial Dirac brackets of the canonical variables
become: 
\begin{eqnarray}
\{\psi ,\Bar{\psi}\}_{D+}^{L} &=&\{\Bar{\psi},\psi \}_{D+}^{L}=-\dfrac{i}{%
\hbar c}\delta ({\boldsymbol{r}}-{\boldsymbol{r}}^{\prime })\ ,  \notag \\
\{\pi _{L},\Bar{\pi}_{L}\}_{D+}^{L} &=&\{\Bar{\pi}_{L},\pi _{L}\}_{D+}^{L}=%
\dfrac{i\hbar c}{4}\delta ({\boldsymbol{r}}-{\boldsymbol{r}}^{\prime })\ , 
\notag \\
\{\psi ,\pi _{L}\}_{D+}^{L} &=&\{\pi _{L},\psi \}_{D+}^{L}=\{\Bar{\psi},\Bar{%
\pi}_{L}\}_{D+}^{L}=\{\Bar{\pi}_{L},\Bar{\psi}\}_{D+}^{L}=-\dfrac{1}{2}%
\delta ({\boldsymbol{r}}-{\boldsymbol{r}}^{\prime })\ ,  \label{DiracL}
\end{eqnarray}%
and%
\begin{eqnarray}
\{\psi ,\Bar{\psi}\}_{D+}^{R} &=&\{\Bar{\psi},\psi \}_{D+}^{R}=\dfrac{i}{%
\hbar c}\delta ({\boldsymbol{r}}-{\boldsymbol{r}}^{\prime })\ ,  \notag \\
\{\pi _{R},\Bar{\pi}_{R}\}_{D+}^{R} &=&\{\Bar{\pi}_{R},\pi _{R}\}_{D+}^{R}=-%
\dfrac{i\hbar c}{4}\delta ({\boldsymbol{r}}-{\boldsymbol{r}}^{\prime })\ , 
\notag \\
\{\psi ,\pi _{R}\}_{D+}^{R} &=&\{\pi _{R},\psi \}_{D+}^{R}=\{\Bar{\psi},\Bar{%
\pi}_{R}\}_{D+}^{R}=\{\Bar{\pi}_{R},\Bar{\psi}\}_{D+}^{R}=-\dfrac{1}{2}%
\delta ({\boldsymbol{r}}-{\boldsymbol{r}}^{\prime })\ .  \label{DiracR}
\end{eqnarray}%
The relations \ref{DiracL} containing left derivatives have also been given
in Ref. \cite{Sundermeyer}, however in different notation (our momenta
differ by factors of $\gamma ^{0}$). Note that now only the Dirac brackets
of the canonical pairs are identical when calculated in terms of left or
right derivatives. Hence, the $L,R$ superscripts may be omitted from both
the Poisson and Dirac brackets of the canonical pairs only. Also note the
appearance of the $1/2$ factors, similarly to the Dirac brackets of the
canonical pairs in the spinor formalism.

\subsection{Reduced phase space of Grassmann variables}

It is easy to identify a canonical chart span by the canonical pair 
\begin{equation}
\left( 
\begin{array}{c}
\psi _{1}^{L}=\frac{1}{2}\psi +\frac{i}{\hbar c}\bar{\pi}_{L} \\ 
\pi _{1}^{L}=-\frac{i\hbar c}{2}\bar{\psi}+\pi _{L}%
\end{array}%
\right) ~  \label{Psi1Pi1L}
\end{equation}%
on the reduced phase space and a second canonical pair 
\begin{equation}
\left( 
\begin{array}{c}
\psi _{2}^{L}=\frac{1}{2}\psi +\frac{i}{\hbar c}\bar{\pi}_{L}=-\dfrac{i}{%
\hbar c}\phi ^{L} \\ 
\pi _{2}^{L}=\frac{i\hbar c}{2}\bar{\psi}+\pi _{L}=\Bar{\phi}^{L}%
\end{array}%
\right) ~  \label{Psi2Pi2L}
\end{equation}%
representing the constraints. We have checked that in terms of this
canonical chart the Dirac bracket (\ref{DL}) on the full phase space becomes
the Poisson bracket (\ref{PL}) on the reduced phase space (thus with
variables $\psi _{1}^{L},\pi _{1}^{L}$ only).

Similarly the canonical pair%
\begin{equation}
\left( 
\begin{array}{c}
\psi _{1}^{R}=\frac{1}{2}\psi -\frac{i}{\hbar c}\bar{\pi}_{R} \\ 
\pi _{1}^{R}=\frac{i\hbar c}{2}\bar{\psi}+\pi _{R}%
\end{array}%
\right)  \label{Psi1Pi1R}
\end{equation}%
on the reduced phase space and the pair of constraints 
\begin{equation}
\left( 
\begin{array}{c}
\psi _{2}^{R}=\frac{1}{2}\psi +\frac{i}{\hbar c}\bar{\pi}_{R}=\dfrac{i}{%
\hbar c}\phi ^{R} \\ 
\pi _{2}^{R}=\pi _{R}-\frac{i\hbar c}{2}\bar{\psi}=\Bar{\phi}^{R}%
\end{array}%
\right)  \label{Psi2Pi2R}
\end{equation}%
generate another canonical chart for which the Dirac bracket (\ref{DR}) on
the full phase space becomes the Poisson bracket (\ref{PR}) on the reduced
phase space.

\section{Canonical Second Quantization of the Dirac field}

We have seen that a classical Hamiltonian constrained treatment of the Dirac
field is possible is several ways. We have shown, how a Poisson structure
can be employed at the price of redefining both the Poisson and Dirac
brackets such that a factor ordering necessary for the Dirac spinor
interpretation of the Dirac field can be retained. Alternatively, regarding
the Dirac field as a Grassman odd variable, the Hamiltonian treatment can
also be pursuit in terms of generalized Poisson and generalized Dirac
structures for anticommuting fields.

Heuristically, the canonical quantization can be achieved from both type of
approaches, as presented below.

\subsection{Spinorial approach}

Due to the factor ordering requirements imposed in the classical discussion
one has to pay special attention to reflect this in the canonical
quantization as well.

A suitable recipe to be applied here is to transform the Dirac bracket of a
Dirac spinor $\mathcal{D}$ (depending on the phase space variables $\psi ,%
\bar{\pi}$) with a Dirac adjoint spinor $\mathcal{A}$ (depending on the
phase space variables $\bar{\psi},\pi $) into an anticommutator as follows 
\begin{equation}
\left\{ \mathcal{D},\mathcal{A}\right\} _{D}\mapsto -\dfrac{i}{\hbar c}[%
\widehat{\mathcal{D}},\widehat{\mathcal{A}}\gamma ^{0}]_{+}~.  \label{canSp}
\end{equation}%
In particular, for the nonvanishing Dirac brackets (\ref{canDBa}), (\ref%
{canDBb}), (\ref{canDB1}), (\ref{canDB2}) the above prescription yields 
\begin{equation}
\left[ \widehat{\psi }\left( t,\mathbf{r}\right) ,\widehat{\psi }^{\dag
}\left( t,\mathbf{r}^{\prime }\right) \right] _{+}=\delta \left( \mathbf{r}-%
\mathbf{r}^{\prime }\right) ~,  \label{Sp1}
\end{equation}%
which is the starting point of the second quantization of the Dirac field,
together with the anticommutators%
\begin{equation}
\left[ \widehat{\bar{\pi}}\left( t,\mathbf{r}\right) ,\widehat{\pi }\left( t,%
\mathbf{r}^{\prime }\right) \gamma ^{0}\right] _{+}=\left( \frac{\hbar c}{2}%
\right) ^{2}\delta \left( \mathbf{r}-\mathbf{r}^{\prime }\right) ~,
\label{Sp2}
\end{equation}%
\begin{equation}
\left[ \widehat{\psi }\left( t,\mathbf{r}\right) ,\widehat{\pi }\left( t,%
\mathbf{r}^{\prime }\right) \gamma ^{0}\right] _{+}=\frac{i\hbar c}{2}\delta
\left( \mathbf{r}-\mathbf{r}^{\prime }\right) ~,  \label{Sp3}
\end{equation}%
\begin{equation}
\left[ \widehat{\bar{\pi}}\left( t,\mathbf{r}\right) ,\widehat{\psi ^{\dag }}%
\left( t,\mathbf{r}^{\prime }\right) \right] _{+}=-\frac{i\hbar c}{2}\delta
\left( \mathbf{r}-\mathbf{r}^{\prime }\right) ~.  \label{Sp4}
\end{equation}%
The constraints (\ref{primary}) give the operator identities%
\begin{equation}
\widehat{\pi }=\frac{i\hbar c}{2}\widehat{\bar{\psi}}~,\quad \widehat{\bar{%
\pi}}=-\frac{i\hbar c}{2}\widehat{\psi }~.  \label{opidSp}
\end{equation}%
Substituting them into Eqs. (\ref{Sp2}), (\ref{Sp3}) and (\ref{Sp4}) all
reproduce Eq. (\ref{Sp1}), which is the fundamental anticommutator between
the operator $\widehat{\psi }$ and its adjoint $\widehat{\psi }^{\dag }$ in
the quantum treatment of the Dirac field \cite{BD,HK,IZ}.

The canonical quantization needs to be defined for arbitrary scalar phase
space functions $f,g$ as well. As they are linear combinations of the type $%
\sum_{i}\mathcal{A}_{i}\mathcal{D}_{i}$ and the Leibniz rule holds%
\begin{eqnarray}
\left\{ f,g\right\} _{D} &=&\sum_{ij}\left\{ \mathcal{A}_{i}^{\left(
f\right) }\mathcal{D}_{i}^{\left( f\right) },\mathcal{A}_{j}^{\left(
g\right) }\mathcal{D}_{j}^{\left( g\right) }\right\} _{D}  \notag \\
&=&\sum_{ij}\mathcal{A}_{i}^{\left( f\right) }\left\{ \mathcal{D}%
_{i}^{\left( f\right) },\mathcal{A}_{j}^{\left( g\right) }\right\} _{D}%
\mathcal{D}_{j}^{\left( g\right) }-\sum_{ij}\mathcal{A}_{j}^{\left( g\right)
}\left\{ \mathcal{D}_{j}^{\left( g\right) },\mathcal{A}_{i}^{\left( f\right)
}\right\} _{D}\mathcal{D}_{i}^{\left( f\right) }~.
\end{eqnarray}%
The factor ordered Dirac bracket terms $\left\{ \mathcal{D}_{i}^{\left(
f\right) },\mathcal{D}_{j}^{\left( g\right) }\right\} _{D}$ and $\left\{ 
\mathcal{A}_{i}^{\left( f\right) },\mathcal{A}_{j}^{\left( g\right)
}\right\} _{D}$ gave no contribution (due to their second and first
arguments, respectively) and we have employed the antisymmetry of the Dirac
bracket in the last term. Then, the canonical quantization recipe (\ref%
{canSp}) can be applied to $\left\{ f,g\right\} _{D}$ as well.

The canonical quantization can also be done by starting from the reduced
phase space variables (\ref{Psi1Pi1}), for which we apply the prescription (%
\ref{canSp}), obtaining%
\begin{equation}
\lbrack \widehat{\psi _{1}},\widehat{\pi _{1}}\gamma ^{0}]_{+}=i\hbar
c\delta \left( \mathbf{r}-\mathbf{r}^{\prime }\right) ~.
\end{equation}%
By employing the operator identities (\ref{opidSp}), we recover the
fundamental anticommutator (\ref{Sp1}) once more.

\subsection{Grassmannian approach with left derivatives}

The canonical quantization recipe in this case is%
\begin{equation}
\left\{ F,G\right\} _{D+}^{L}\mapsto -\dfrac{i}{\hbar c}[\widehat{F},%
\widehat{G}\gamma ^{0}]_{+}~.
\end{equation}%
After the changes in notation, this quantization scheme is equivalent to the
one proposed by Casalbuoni \cite{Casalbuoni}.

With the operator identities%
\begin{equation}
\widehat{\Bar{\pi}}_{L}=-\dfrac{i\hbar c}{2}\widehat{\psi }~,\quad \widehat{%
\pi }_{L}=-\dfrac{i\hbar c}{2}\widehat{\Bar{\psi}}~.
\end{equation}%
arising from the constraints (\ref{primaryL}), all quantized versions of the
nontrivial Dirac brackets (\ref{DiracL}) reproduce the fundamental
anticommutator (\ref{Sp1}).

\subsection{Grassmannian approach with right derivatives}

In this case, the canonical quantization recipe becomes 
\begin{equation}
\left\{ F,G\right\} _{D+}^{R}\mapsto \dfrac{i}{\hbar c}[\widehat{F},\widehat{%
G}\gamma ^{0}]_{+}~.  \label{anticomR}
\end{equation}%
The constrains (\ref{primaryR}) lead to the operator identities 
\begin{equation}
\widehat{\Bar{\pi}}_{R}=\dfrac{i\hbar c}{2}\widehat{\psi }~,\quad \widehat{%
\pi }_{R}=\dfrac{i\hbar c}{2}\widehat{\Bar{\psi}}~.  \label{piR}
\end{equation}%
With these, the fundamental anticommutator (\ref{Sp1}) is once again
reproduced from all nontrivial Dirac brackets (\ref{DiracR}).

\section{On the coupled gravity-Dirac field system}

Instead of a metric description, suitable to deal with bosonic (tensorial)
fields, gravity can also be described in terms of 16 tetrad variables, as
worked out in 1929 by Weyl \cite{Weyl29}. The four tetrad vectors (with 4
spacetime components each) represent local orthonormal Lorentz frames at
each space-time point with respect to which spinors can be defined. This
choice of gravitational variables naturally allows for covariant derivatives
of spinors, in which the spin connection (or Ricci rotation coefficients,
with 24 components, having two tetrad indices) couples to the $\gamma ^{a}$
matrices.

The canonical (Arnowitt---Deser---Misner type) formulation of the dynamics
of gravity coupled to the Dirac field has been studied long time ago. As
early as 1963 Kibble discussed the canonical form of the Lagrangian
describing the interacting system of gravity and Dirac fields, in the
framework of the tetrad formalism of gravity \cite{Kibble63}. He derived the
dynamics of the Dirac field from a sum of the simplest Lagrangian density $%
\mathfrak{L}_{BD}$ with a spin--spin interaction Lagrangian (bearing
similarity to the one arising in the Fermi interaction). Our analysis, by
contrast, proceeds from the Hermitian Lagrangian density $\mathfrak{L}_{IZ}$%
. Kibble assumed anticommuting Dirac spinors, similarly as in our discussion
employing anticommuting Grassmann variables. In the first part of our
treatment however, we introduced factor ordering instead (requiring Dirac
adjoint spinors to multiply Dirac spinors, in this order).

Deser and Isham in Ref. \cite{DeserIsham76} provided a crystal-clear
canonical description of the gravitational sector in terms of tetrad
variables, in which the role of the six pairs of geometrodynamical variables
(the induced metric components and associated momenta, the trace-modified
extrinsic curvature) is taken by the twelve pairs of variables related to
the spatial tetrad components and their respective associated momenta. While
the six pairs of geometrodynamical variables can be reduced to two pairs (as
the first class Hamiltonian and diffeomorphism constraints each reduce the
degrees of freedom by two \cite{Sundermeyer}), in the tetrad formalism
additional degrees of freedom emerge due to the six-parameter Lorentz
transformations of the tetrad. Each of the related additional six angular
momentum type constraints, being also first class, reduce the degrees of
freedom by two, reestablishing two pairs of canonical variables as physical.
The Lorentz invariance of the gravitational action was re-established by
inserting the angular momentum type constraints into the action with a set
of six Lagrange multipliers. Although the canonical Dirac action for a
coupled Dirac field has not been given explicitly, no other obstacle for
that was envisioned than the tedious nature of the involved calculations.

Both the Einstein and Dirac Lagrangians, also the respective equations of
motion were given in terms of a suitably chosen spinor Lagrangian in tetrads
suited to the 3+1 formalism of general relativity \cite{GeheniauHenneaux77},
reproducing and generalizing earlier results by Dirac \cite{DiracGS}. There,
the spinor Lagrangian is the generalization of $\mathfrak{L}_{IZ}$ to curved
spacetime, however the Hamiltonian given in the Note Added in Proof does not
reproduce our result.

In Ref. \cite{NelsonTeitelboim78} Nelson and Teitelboim gave the action for
interacting gravitational and Dirac fields expressed in canonical form, in
tetrad and spinorial variables. The results of Ref. \cite{DeserIsham76} were
reproduced, supplementing the Hamiltonian and diffeomorphism constraints
with the contributions from the Dirac spinor. The emerging Dirac Hamiltonian
in the flat limit $L^{i}\mapsto \gamma ^{i}$ does not agree with our
Hamiltonian (\ref{HIZ}). Agreement can be reached only at the price of
inserting the definitions of the momenta in the Hamiltonian (a forbidden
step before variation), and by dropping a spurious factor $i$ from the mass
term of the Hamiltonian given in Ref. \cite{NelsonTeitelboim78}. The Dirac
action in Ref. \cite{NelsonTeitelboim78} is again the covariant version of $%
\mathfrak{L}_{IZ}$ (after omitting a spurious factor $i$ from the mass
term), however the authors do not allow the variation with respect to the
Dirac adjoint spinor. The discussion then proceeded in terms of the real and
imaginary parts of the spinor field, hence the comparison with our methods
is not immediate. Solely the brackets of the spinor field and its Dirac
adjoint are given, reproducing our Dirac bracket (\ref{canDBa}) in the flat
limit, however the Dirac brackets (\ref{canDB1}-\ref{canDB2}) between the
canonical variables, which are the most important in our discussion, are
missing.

Therefore, our discussion goes beyond the flat spacetime limit of the
results of the references mentioned above.

It is well known that gravity can be also described in terms of torsion \cite%
{Hayashi, Hayashi2} or nonmetricity \cite{NewerGR}, recently summarized in
Ref. \cite{Heinsenberg1}. The three approaches are sometimes referred as the
geometric trinity of gravity \cite{Trinity}. It has been proven by Koivisto,
that the Dirac fermions described by the covariant version of the Hermitian
Lagrangian $\mathcal{L_{IZ}}$ are unaffected by non-metricity \cite{Koivisto}%
. Moreover, the consistency of the coupling to the torsion has been disputed 
\cite{Obukhov}. Hence, when the connection does not have a contorsion part,
but it has a disformation part (related to its nonmetricity $Q_{abc}$),
Dirac fermions do not couple to the latter and their minimal coupling to any
torsionfree gravity theory is viable \cite{HeisenbergJCAP, telecoupling}.

Modified gravity generalizations of both $f(R)$ (see the monography \cite%
{HarkoLobo} with references therein, also \cite{LeviSaid}) and $f(Q)$ types
(see Refs. \cite{LazkozQ, BoehmerQ, HeisenbergQ}) have been discussed
extensively. Dirac fields in $f(Q)$ theories were investigated in Ref. \cite%
{CarloniQ}. In principle, $f(R,Q)$ type gravity models could also properly
incorporate minimally coupled Dirac spinors.

\section{Concluding Remarks}

We generalized a previously successful discussion of the Schr\"{o}dinger
field \cite{Schrod} as a constrained system through the Dirac--Bergmann
algorithm to the case of the Dirac field. This was motivated by the analogy
of the Lagrangians describing the Dirac field with the simplest Schiff and
Hermitian Henley--Thirring Lagrangians of the Schr\"{o}dinger field.

In order to follow as closely as possible the treatment of the Schr\"{o}%
dinger field, we discussed the Dirac field classically as a spinor. Then,
for the Hamiltonian evolution we introduced suitably modified, factor
ordered Poisson and Dirac brackets (in which Dirac adjoint spinors are
always followed by spinors) and we recovered all essential ingredients,
similarly to the case of the Schr\"{o}dinger field. We introduced momenta,
which are Dirac adjoints of the respective spinorial field variables. This
ensured that no $\gamma ^{0}$ factors appear in the Poisson bracket.

The Dirac--Bergmann algorithm applied to the Hermitian Lagrangian for the
Dirac field gave two primary Hamiltonian constraints, Dirac adjoints of each
other. Evolving (through the factor ordered Poisson bracket) the constraints
generated the Dirac equation\ and its Dirac adjoint as consistency
conditions, holding weakly. The constraints being second class implied the
introduction of the factor ordered Dirac bracket, generating time evolution.
The Dirac brackets of the canonical pairs contained $1/2$ factors not
present in their Poisson brackets, similarly to the case of the Schr\"{o}%
dinger field, an essential feature in its canonical quantization.

We identified the canonical chart in which one canonical pair spans the
reduced phase space while the other is the pair of constraints. In this
chart the Dirac bracket on the full phase space became the Poisson bracket
on the reduced phase space, on which\ the Hamiltonian coincides with the
Hermitian Hamiltonian evaluated on shell. This feature differs from the Schr%
\"{o}dinger case, where the reduced Hamiltonian coincides on shell with both
the Schiff and Henley--Thirring Hamiltonians. For the Dirac field the
Hamiltonian obtained from the simplest Lagrangian (the analogue of the
Schiff Hamiltonian) carries additional contributions. The Dirac equation and
its Dirac adjoint could be recovered as canonical equations on the reduced
phase space.

When the Dirac field is considered an odd Grassmann variable, it obeys the
same algebraic structure classically as in the quantum version of the
theory. Hence, we carried out the Dirac--Bergmann algorithm for the
Grassmannian Dirac field with both left and right derivatives acting on
Grassmann valued superfunctions. These involved other types of generalized
Poisson and Dirac brackets, symmetric in the canonical data, rather than
antisymmetric.

Finally, we provided the recipe for the canonical second quantization of all
three versions of the classical Dirac dynamics. This naturally transformed
all three versions of the introduced Dirac brackets into the fundamental
anticommutator characterizing the second quantized Dirac field.

\appendix*

\section{The Dirac spinor represents spin $1/2$ particles}

Taking any spacetime vector $V^{a}$, one can define another vector with
identical components on the space spanned by the orthonormal basis $\left\{ 
\mathrm{I}_{2},\sigma ^{\mu }\right\} $: 
\begin{eqnarray}
v &=&V^{0}\left( 
\begin{array}{cc}
1 & 0 \\ 
0 & 1%
\end{array}%
\right) +V^{1}\left( 
\begin{array}{cc}
0 & 1 \\ 
1 & 0%
\end{array}%
\right) +V^{2}\left( 
\begin{array}{cc}
0 & -i \\ 
i & 0%
\end{array}%
\right) +V^{3}\left( 
\begin{array}{cc}
1 & 0 \\ 
0 & -1%
\end{array}%
\right)  \notag \\
&=&\left( 
\begin{array}{cc}
V^{0}+V^{3} & V^{1}-iV^{2} \\ 
V^{1}+iV^{2} & V^{0}-V^{3}%
\end{array}%
\right) ~.
\end{eqnarray}%
A rotation 
\begin{equation*}
\left( 
\begin{array}{c}
V^{0\prime } \\ 
V^{1\prime } \\ 
V^{2\prime } \\ 
V^{3\prime }%
\end{array}%
\right) =\left( 
\begin{array}{cccc}
1 & 0 & 0 & 0 \\ 
0 & \cos \varphi & \sin \varphi & 0 \\ 
0 & -\sin \varphi & \cos \varphi & 0 \\ 
0 & 0 & 0 & 1%
\end{array}%
\right) \left( 
\begin{array}{c}
V^{0} \\ 
V^{1} \\ 
V^{2} \\ 
V^{3}%
\end{array}%
\right)
\end{equation*}%
with angle $\varphi $ about the $z$-axis generates%
\begin{equation}
v^{\prime }=\left( 
\begin{array}{cc}
V^{0}+V^{3} & \left( V^{1}-iV^{2}\right) e^{i\varphi } \\ 
\left( V^{1}+iV^{2}\right) e^{-i\varphi } & V^{0}-V^{3}%
\end{array}%
\right) ~,
\end{equation}%
thus an $SL\left( 2,%
\mathbb{C}
\right) $ transformation on $v^{a}$%
\begin{equation}
v^{\prime }=\left( 
\begin{array}{cc}
e^{i\varphi /2} & 0 \\ 
0 & e^{-i\varphi /2}%
\end{array}%
\right) v\left( 
\begin{array}{cc}
e^{i\varphi /2} & 0 \\ 
0 & e^{-i\varphi /2}%
\end{array}%
\right) ^{\dag }~.
\end{equation}

It is simple to check that $\det v=0$ is equivalent with $V^{a}$ being null
and we choose $V^{a}=\frac{1}{2}\left( 1,0,0,1\right) $ in order to simplify
the forthcoming results. Then, 
\begin{equation}
v=\left( 
\begin{array}{cc}
1 & 0 \\ 
0 & 0%
\end{array}%
\right) =\left( 
\begin{array}{c}
1 \\ 
0%
\end{array}%
\right) \left( 
\begin{array}{cc}
1 & 0%
\end{array}%
\right) ~.
\end{equation}%
The first factor is a left Weyl spinor $\psi _{L}$, while the second is the
transposed of the Weyl adjoint $\bar{\psi}_{R}=\varepsilon \psi _{R}^{\dag }$
of a right Weyl spinor $\psi _{R}$ \cite{Steane}, where 
\begin{equation}
\varepsilon =\left( 
\begin{array}{cc}
0 & 1 \\ 
-1 & 0%
\end{array}%
\right) ~.
\end{equation}%
Under rotations they transform as%
\begin{eqnarray}
\psi _{L} &=&\left( 
\begin{array}{c}
1 \\ 
0%
\end{array}%
\right) \rightarrow \psi _{L}^{\prime }=\left( 
\begin{array}{cc}
e^{i\varphi /2} & 0 \\ 
0 & e^{-i\varphi /2}%
\end{array}%
\right) \left( 
\begin{array}{c}
1 \\ 
0%
\end{array}%
\right) =e^{i\varphi /2}\left( 
\begin{array}{c}
1 \\ 
0%
\end{array}%
\right) ~,  \notag \\
\bar{\psi}_{R}^{T} &=&\left( 
\begin{array}{cc}
1 & 0%
\end{array}%
\right) \rightarrow \bar{\psi}_{R}^{\prime T}=\left( 
\begin{array}{cc}
1 & 0%
\end{array}%
\right) \left( 
\begin{array}{cc}
e^{i\varphi /2} & 0 \\ 
0 & e^{-i\varphi /2}%
\end{array}%
\right) ^{\dag }=e^{-i\varphi /2}\left( 
\begin{array}{cc}
1 & 0%
\end{array}%
\right) ~.
\end{eqnarray}%
From the latter%
\begin{equation}
\psi _{R}=\left( 
\begin{array}{cc}
0 & 1%
\end{array}%
\right) \rightarrow \psi _{R}^{\prime }=e^{i\varphi /2}\left( 
\begin{array}{cc}
0 & 1%
\end{array}%
\right) ~
\end{equation}%
can be easily deduced.

Hence, the Dirac spinor, defined through 
\begin{equation}
\psi =\left( 
\begin{array}{c}
\psi ^{L} \\ 
\psi _{R}^{T}%
\end{array}%
\right)
\end{equation}%
transforms as 
\begin{equation}
\psi ^{\prime }=e^{i\varphi /2}\psi ~.
\end{equation}%
This is simple proof that the Dirac spinor represents spin 1/2 particles.

Note the similarity of this proof to the electromagnetic and gravitational
cases. For the electromagnetic plane wave 
\begin{equation}
A_{a}=\mathfrak{e}_{a}\exp \left( ik^{c}x_{c}\right) +\text{c.c.}
\end{equation}%
(with null wave vector $k^{c}$) the polarization vector $\mathfrak{e}_{a}$
in the Lorenz gauge ensuring transversality and the cancellation of the
temporal and longitudinal polarizations is $\mathfrak{e}_{a}=\left( 0,%
\mathfrak{e}_{1},\mathfrak{e}_{2},0\right) $. Under rotations about the $z$%
-axis with angle $\varphi $ the polarizations $\mathfrak{e}_{\pm }=\mathfrak{%
e}_{1}\mp i\mathfrak{e}_{2}$ transform as $\mathfrak{e}_{\pm }^{\prime
}=e^{\pm i\varphi }\mathfrak{e}_{\pm }$, implying helicity $\pm 1$ for the
photon.

A similar reasoning stands for the gravitational waves in the weak field
approach. For the perturbation $h_{ab}$ of the Minkowski metric, assumed as
a plane wave%
\begin{equation}
h_{ab}=\mathfrak{e}_{ab}\exp \left( ik^{c}x_{c}\right) +\text{c.c.}
\end{equation}%
the polarization tensor $\mathfrak{e}_{ab}$ in harmonic coordinates and
transverse traceless gauge is 
\begin{equation}
\mathfrak{e}_{ab}=\left( 
\begin{array}{cccc}
0 & 0 & 0 & 0 \\ 
0 & \mathfrak{e}_{11} & \mathfrak{e}_{12} & 0 \\ 
0 & \mathfrak{e}_{12} & -\mathfrak{e}_{11} & 0 \\ 
0 & 0 & 0 & 0%
\end{array}%
\right) ~.
\end{equation}%
Under rotations about the $z$-axis with angle $\varphi $ the polarizations $%
\mathfrak{e}_{\pm }=\mathfrak{e}_{11}\mp i\mathfrak{e}_{12}$ transform as $%
\mathfrak{e}_{\pm }^{\prime }=e^{\pm 2i\varphi }\mathfrak{e}_{\pm }$,
signifying helicity $\pm 2$ for the gravitational wave.

\end{document}